\pdfoutput=1
\documentclass[a4paper,english,nolineno]{socg-lipics-v2019} 

\usepackage{url}

\newcommand{\remove}[1]{}  

\usepackage{myMac-lipics}

\remove{ 
\EventEditors{Gill Barequet and Yusu Wang}
\EventNoEds{2}
\EventLongTitle{35th International Symposium on Computational Geometry (SoCG 2019)}
\EventShortTitle{SoCG 2019}
\EventAcronym{SoCG}
\EventYear{2019}
\EventDate{June 18--21, 2019}
\EventLocation{Portland, Oregon, USA}
\EventLogo{socg-logo}
\SeriesVolume{129}
\ArticleNo{43} 
} 

	\usepackage{graphics}
	\usepackage{epsfig}
	
	\newcommand{\nopath}{{\tt NO-PATH}}

	\newcommand{\Elem}{\ensuremath{\mathcal E}}

	\newcommand{\TTT}{{\cal T}}

	\newcommand{\Cl}{\mathop{C\ell}}	
	\newcommand{\cspacehat}{\ensuremath{\wh{C}_{space}}}
	\newcommand{\whphi}{\wh{\phi}}

	\newcommand{\full}{\ensuremath{\mathtt{FULL}}}
	\newcommand{\emptY}{\ensuremath{\mathtt{EMPTY}}}
	\newcommand{\splittt}{\ensuremath{\mathtt Split}}

	\newcommand{\whq}{\wh{q}}	
	\newcommand{\cspace}{\ensuremath{C_{space}}}
	\newcommand{\cfree}{\ensuremath{C_{free}}}

	\newcommand{\getnext}{{\tt GetNext}}

	\newcommand{\sqs}{{\wh{S^2}}}

	\newcommand{\FP}{\ensuremath{F\!p}}
	\newcommand{\Fp}{\ensuremath{F\!p}}
	\newcommand{\wtFp}{\ensuremath{\wt{F\!p}}}

	\newcommand{\free}{\ensuremath{\mathtt{FREE}}}
	
	\newcommand{\stuck}{\ensuremath{\mathtt{STUCK}}}
	\newcommand{\mixed}{\ensuremath{\mathtt{MIXED}}}
	
	\newcommand{\wtC}{\wt{C}}
	\newcommand{\wtphi}{\wt{\phi}}

\title{Rods and Rings: Soft Subdivision Planner
            for $\RR^3\times S^2$
\footnote{The conference version of this paper will appear in
{\it Proc.\ Symposium on Computational Geometry (SoCG '19)}, June, 2019. }
      }  
\titlerunning{Rods and Rings Planner} 

\author{Ching-Hsiang Hsu}
{Department of Computer Science, Courant Institute, New York University, 
New York, NY, USA}
{chhsu@nyu.edu} 
{} 
{Supported by NSF Grant \#CCF-1423228} 
\author{Yi-Jen Chiang}
{Department of Computer Science and Engineering, Tandon School of Engineering,
New York University, Brooklyn, NY, USA}
{chiang@nyu.edu}
{}
{}
\author{Chee Yap}
{Department of Computer Science, Courant Institute, 
New York University, New York, NY, USA}
{yap@cs.nyu.edu} 
{} 
{Supported in part by NSF Grants \#CCF-1423228 and \#CCF-1563942.}

\authorrunning{Hsu, Chiang, and Yap} 

\Copyright{
Ching-Hsiang Hsu and
Yi-Jen Chiang and
Chee Yap
}


\supplement{}

\funding{
Supported in part by NSF Grants \#CCF-1423228 and \#CCF-1563942.
}

\acknowledgements{} 

\begin{document}
\maketitle

\begin{abstract}
	We consider path planning for a rigid spatial robot
	moving amidst polyhedral obstacles.
	Our robot is either a rod or a ring.
	Being axially-symmetric, their configuration space 
	is $\RR^3\times S^2$ with 5 degrees of freedom (DOF).
	Correct, complete and practical path planning for such robots
	is a long standing challenge in robotics.
	While the rod is one of the most widely studied
	spatial robots in path planning,
	the ring seems to be new, and a rare example
	of a non-simply-connected robot.
	This work provides rigorous and
	complete algorithms for these robots
        with theoretical guarantees.
	We implemented the algorithms
	in our open-source Core Library.  Experiments
	show that they are practical, achieving
	near real-time performance.  We compared our planner
	to state-of-the-art sampling planners in OMPL~\cite{ompl}.
	
	Our subdivision path planner is based on 
	the twin foundations of $\vareps$-exactness and soft predicates.
	Correct implementation is relatively easy.
	The technical innovations include
	subdivision atlases for $S^2$, introduction of $\Sigma_2$
	representations for footprints, and
	extensions of our feature-based technique for ``opening up the
	blackbox of collision detection''.
	
	\ignore{
	  we device a splitting strategy to achieve real-time performance.
	  We implemented this planner, and our experiments on a variety
	  of challenging obstacle environments confirms its practicality.
	  Unlike most practical planners today, we do not use randomization
	  and yet offer much stronger theoretical guarantees of performance.
	
	  Moreover, a variety of very interesting 5-DOF robots are now
	  practical:
	  disc (flying saucer), cone (bullet or rocket), or a ring.
	}%
\end{abstract}
%

\sect[1]{Introduction}
	Motion planning 
	\cite{lavalle:planning:bk,choset-etal:bk}
	is a fundamental topic in robotics because the typical robot
	is capable of movement.  Such algorithms 
	are increasingly relevant with the
	current surge of interest in inexpensive commercial mobile robots,
	from domestic robots that vacuum the floor to drones that
	deliver packages.
	We focus on what is called \dt{path planning} which, in its 
	elemental form, asks for a collision-free path from a start to a goal
	position, assuming a known environment.  Path planning is based on
	robot kinematics and collision-detection only, and the variety
	of such problems are surveyed in
	\cite{hss:motionplanning:crc:17}.
	The output of a ``path planner'' is either a path or a \nopath,
	signifying that no path exists.
	Remarkably, the single bit of information encoded by
	\nopath\ is often missing in discussions. 
	The standard definitions of correctness for path planners
	(\dt{resolution completeness}
	and \dt{probabilistic completeness})
	omit this bit \cite{wang-chiang-yap:motion-planning:15}.
	The last 30 years have seen a flowering of practical path planning
	algorithms.  The dominant algorithmic paradigm of these planners
	has been variants of the \dt{Sampling Approach} such as PRM, EST, RRT,
	SRT, etc (see \cite[p.~201]{choset-etal:bk}).
	Because this bit of information is not built into the specification
	of such algorithms, it has led to non-termination issues and
	a large literature addressing the ``narrow passage problem''
	(e.g., \cite{nowakiewicz:mst:10,lazy-toggle-prm:13}).
	\ignore{
	Quoting the chapter on PRM-like methods \cite[p.201]{choset-etal:bk}:
	``{\em PRM, EST, RRT, SRT, and their variants have changed the way path
	planning is performed for high-dimensional robots. They have also paved
	the way for the development of planners for problems beyond basic path
	planning.}''
	}
	Our present paper is based on the \dt{Subdivision Approach}.
	This approach has a venerable history in robotics -- see 
	\cite{brooks-perez:subdivision:83,zhu-latombe:hierarchical:91}
	for early planners based on subdivision.

	\ignore{%
	The term \dt{motion planning} is very general:
	The solution to any motion planning problem is a \dt{motion}.
	Underlying the motion is a path in configuration space,
	called its \dt{(kinematic) path}.  Since we only focus on this
	path, our problem is commonly known as
	``path planning'' and the corresponding algorithms as ``planners''.
	Motion planning addresses many other
	issues beyond kinematics and collision-detection:
	call them \dt{extra-kinematic} issues.
	E.g., in trajectory planning, we need to
	find a kinematic path together with timing information along the path.
	The timing information may have to satisfy various
	constraints based on velocity, acceleration, dynamics,
	smoothness, optimality, non-holonomic constraints, etc.
	Such problems have very high
	computational complexity or have no exact solutions.
	A pragmatic approach for addressing them is to
	first solve the path planning problem (since \nopath\ implies
	no solution for the extra-kinematic planning problem).
	If a path is found, we then use it as a basis for
	constructing motions that address the extra-kinematic issues.
	It is therefore critical to have a rigorous
	foundation for path finding.
	}%
	Exact path planning has many issues 
	including a serious gap between theory and implementability.
	In \cite{wang-chiang-yap:motion-planning:15,sss}, we
	introduced a theoretical framework based on subdivision
	to close this gap.  This paper demonstrates for the first
	time that our framework
	is able to achieve rigorous state-of-the-art planners in 3D.
	\refFig{rod-trace-subdiv-rand100} shows our rod robot in an
	environment with 100 random tetrahedra.
        \refFig{ring-trace-sub-posts} shows our ring robot in an
	environment with pillars and L-shaped posts.
%
%
	See a video demo from
	\myHrefx{http://cs.nyu.edu/exact/gallery/rod-ring/rod\_ring.html}.

\FigEPS{rod-trace-subdiv-rand100}
		  { 0.25 } 
		  {Rod robot amidst 100 random tetrahedra: 
                   (a) trace of a found path; 
                   (b) subdivision of translational boxes on the path.
%
                  }
\remove{ 
\FigEPS{ring-trace-subdiv}
		  { 0.35 } 
		  {Ring robot amidst two posts:
                   (a) trace of a found path; 
                   (b) subdivision of translational boxes on the path.
                  }
} 
\remove{
	\FigEPS{rod-trace-rand100}
		  { 0.35 } 
		  { TEMP1: Trace of a found path }
	\FigEPS{rod-subdiv-rand100}
		  { 0.35 } 
		  { TEMP1: Subdivision of translational boxes on path} 
	\FigEPS{ring-trace}
		  { 0.35 } 
		  { TEMP2: Trace of a found path }
	\FigEPS{ring-subdiv}
		  { 0.35 } 
		  { TEMP2: Subdivision of translational boxes on path} 
}
\ignore{
	\myfigtwo[ Rod robot amidst 100 random tetrahedra]
		{rod-rand100-gui}
		{rod-trace-rand100}
		  { Trace of a found path }
		    { 0.35 } 
		{rod-subdiv-rand100}
		  { Subdivision of translational boxes on path} 
		    { 0.35 } 
	\myfigtwo[ Ring robot amidst two posts]{ring-2posts-gui}
		{ring-trace}
		  { Trace of a found path }
		    { 0.35 } 
		{ring-subdiv}
		  { Subdivision of translational boxes on path} 
		    { 0.35 } 
}%

	In this paper, we consider a rigid spatial robot $R_0$
	that has an axis of symmetry.
	See \refFig{cone-disc-ring}(a) for several possibilities for $R_0$:
	rod (``ladder''), cone (``space shuttle''), disc
	(``frisbee'') and ring (``space station''). 
	Our techniques easily allow these robots to be ``thickened''
	by Minkowski sum with balls (see~\cite{yap-luo-hsu:thicklink:16}).
	The \dt{configuration space} may be taken to be
	$\cspace=\RR^3\times S^2$ where $S^2$ is the unit $2$-sphere.
	We identify $R_0$ with a closed subset of $\RR^3$,
	called its ``canonical footprint''.
	E.g., if $R_0$ is a rod (resp., ring),
	then the canonical footprint is a line segment (resp., circle)
	in $\RR^3$.  Each configuration $\gamma\in\cspace$ corresponds
	to a rotated translated copy of the canonical footprint,
	which we denote by $\Fp(\gamma)$.  
	Path planning involves another input, the
	\dt{obstacle set} $\Omega\ib\RR^3$ that the robot must avoid.
	We assume that $\Omega$ is a closed polyhedral set.
	Say $\gamma$ is \dt{free} if $\Fp(\gamma)\cap\Omega$ is empty.
	The \dt{free space} comprising all the free configurations
	is an open set by our assumptions, and is
	denoted $\cfree=\cfree(\Omega)$.
	A parametrized continuous curve $\mu:[0,1]\to\cspace$ is
	called a \dt{path} if the range of $\mu$ is in $\cfree$.
	Path planning amounts to finding such paths.
	Following \cite{zhu-latombe:hierarchical:91},
	we need to classify boxes $B\ib\cspace$ into one of three types:
	\free, \stuck\ or \mixed.  Let $C(B)$ denote the classification
	of $B$: $C(B)=\free$ if $B\ib\cfree$, and
	$C(B)=\stuck$ if $B$ is in the interior of $\cspace\setminus\cfree$.
	Otherwise, $C(B)=\mixed$.
	%
	\ignore{
	But we need to generalize it to the classification of
	an entire set $B\ib\cspace$.
	Define the \dt{classification function}
	$C:\cspace\to \set{\free,\stuck,\mixed}$ where $C(\gamma)=\mixed$
	if $\gamma$ is on the boundary of $\cfree$; otherwise,
	$C(\gamma)=\free$ iff $\gamma\in\cfree$.
	For subdivision, we generalize the
	collision-detection predicate to an entire box $B\ib\cspace$.
	Alternatively, we may say \emptY/\full/\mixed\
	as in \cite{zhang-kim-manocha:path-non-existence:08}.
	This is an exact predicate, where \free\ and \stuck\ 
	corresponds to $B\ib\cfree$ and $B\cap\cfree=\es$,
	and otherwise $C(B)=\mixed$.
	}
	One of our goals is to introduce classifications
	$\wtC(B)$ that are ``soft versions'' of $C(B)$ (see Appendix~\ref{app-SSS}). 

%

	    	\begin{figure}[htb]
	    	  \begin{center}
		   \scalebox{0.26}{
	    	     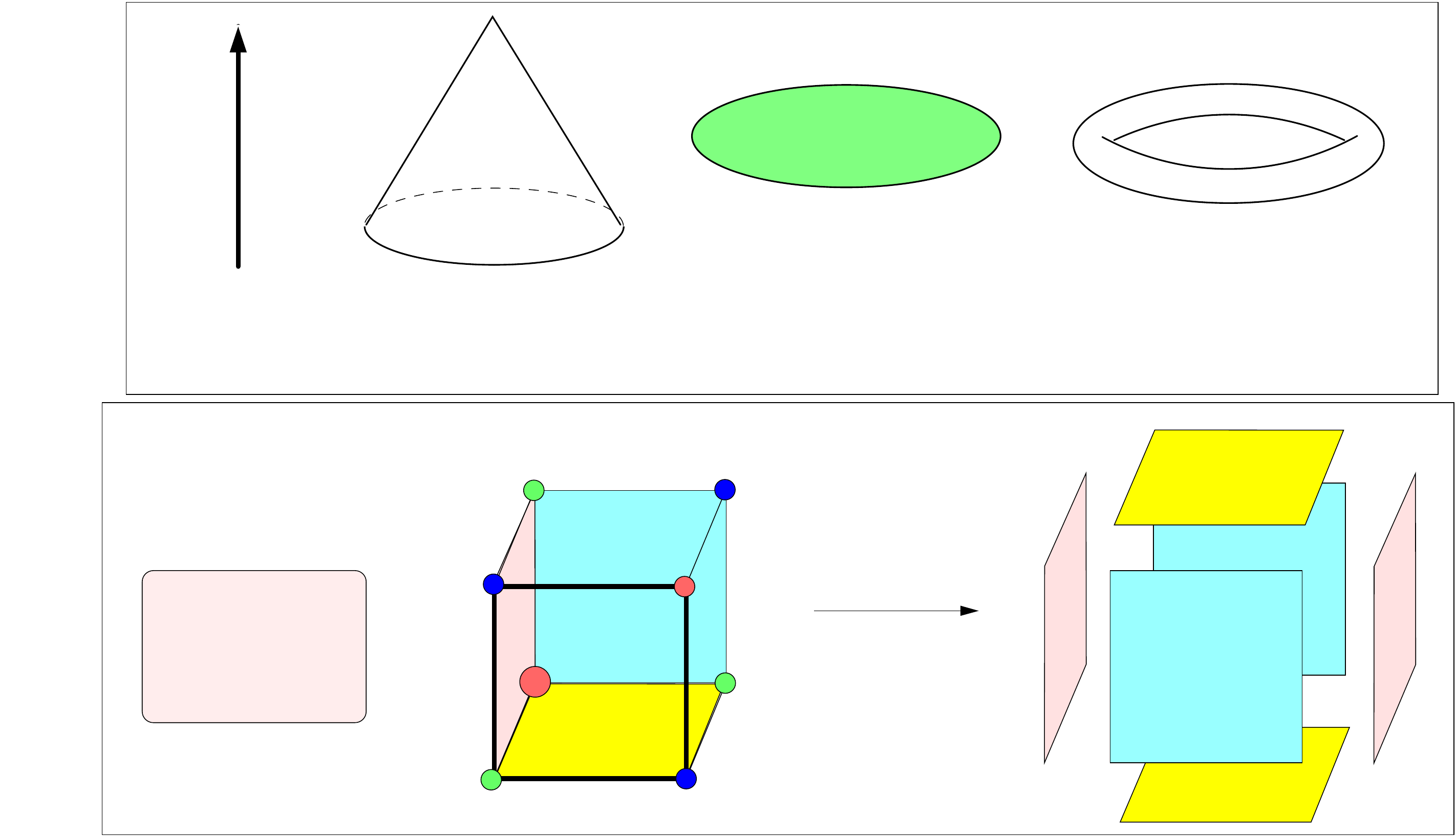}
	    	   \caption{(a) 3D rigid robots with
		      $\cspace=\RR^3\times S^2$.
		      (b) Subdivision atlas for $S^2$ via $\sqs$.
%
}
	    	   \label{fig:cone-disc-ring}
	    	  \end{center}
	    	\end{figure} 	

\remove{
	    	\begin{figure}[htb]
	    	  \begin{center}
		   \scalebox{0.3}{
	    	     \input{figs/cone-disc-ring.pdf_t}}
	    	   \caption{(a) 3D rigid robots with
		      $\cspace=\RR^3\times S^2$.
		      (b) Subdivision atlas for $S^2$ via $\sqs$.
}
	    	   \label{fig:cone-disc-ring}
	    	  \end{center}
	    	\end{figure} 
} 

	We present four desiderata in path planning: \\
{\bf (G0)} the planner must be mathematically rigorous and complete;
\\{\bf (G1)} it must have correct implementations which are also:
\\{\bf (G2)} relatively easy to achieve and
\\{\bf (G3)} practically efficient. \\
	In (G0), we use the standard Computer Science notion of
	an algorithm being \dt{complete} if
	(a) it is partially complete\footnote{
	 	Partial completeness means the algorithm produces
		a correct output {\em provided} it halts.
	}
	and (b) it halts. 
	The notions of
	\dt{resolution completeness} and \dt{probabilistic completeness}
	in robotics have requirement (a) but not (b).
	In probabilistic-complete algorithms,
	halting with \nopath\ is achieved heuristically
	by putting limits on time and/or number of samples.
	But such limits are not intrinsic to the input instance.
	In resolution-complete algorithms, \nopath\ halting is
	based on width $w$ of subdivision box being small enough
	(say $w<\vareps$).  
	One issue is that the width of a box is a direct measure
	of clearance (but there is a nontrivial correlation);
	secondly,
	box predicates are numerical and ``accurate enough''
	($\sigma$-effective in our theory). 
	These issues are exacerbated
	when algorithms do not use box predicates,
	but perform sampling at grid points of the subdivision.
	In contrast, our \nopath\ guarantees an intrinsic property:
	there is no path of clearance $K \vareps$ (see below).

	But desideratum (G0) is only the base line.
        %
          A 
        (G0)-planner may not be worth much in
	a practical area like robotics unless it also 
	has implementations with properties (G1-G3).
	E.g., the usual exact algorithms satisfy (G0)
	but their typical implementations fail (G1).
	With proper methods \cite{sharma-yap:crc},
	it is possible to satisfy (G1); Halperin et al
	\cite{halperin-fogel-wein:bk} give such
	solutions in 2D using 
        \cgal.
%
%
%
	Both (G0) and (G1) can be formalized (see next),
	but (G2) and (G3) are informal.  
        The robotics
	community has developed various criteria to evaluate (G2) and (G3).
	The accepted practice is 
	having an implementation (proving (G2)) that achieves ``real time''
	performance on a suite of non-trivial instances (proving (G3)).

	The main contribution of this paper is the design
	of planners
	for spatial robots with 5 DOFs 
	that have the ``good'' properties (G0-G3).   
	This seems to be the first for such robots.
	To achieve our results, we introduce
	theoretical innovations and algorithmic techniques 
	that may prove to be more widely applicable.

	In path planning and in Computational Geometry,
	there is a widely accepted interpretation of desideratum (G0):
	it is usually simply called ``exact algorithms''.
	But to stress our interest in alternative notions of exactness, we
	refer to the standard notion as \dt{exact (unqualified)}.
	Planners that are exact (unqualified) are first
	shown in \cite{ss2}; this can be viewed as a fundamental
	result on decidability of connectivity in semi-algebraic sets
	\cite{basu-pollack-roy:bk}. 
	The curse of exact (unqualified) algorithms is that the algorithm must 
	detect any degeneracies in the input and
	handle them explicitly.
	But exact (unqualified) algorithms are rare,
	mainly because degeneracies are numerous and hard to analyze:
	the usual expedient is to assume ``nice'' (non-degenerate) inputs.
	So the typical exact (unqualified) algorithms
	in the literature are \dt{conditional} algorithms, i.e.,
	its correctness is conditioned on niceness assumptions.
	Such gaps in exact (unqualified) algorithms are not
	an issue as long as they are not implemented.
	For non-linear problems beyond 2D,
	complete degeneracy analysis is largely non-existent.
	This is vividly seen in the fact that, despite long-time
	interest, there is still no exact (unqualified)
	algorithm for the Euclidean Voronoi diagram of a polyhedral set
	(see \cite{hemmer-setter-halperin:vordiag-3dlines:10,everett+3:vor:09,everett+4:vor:09,yap-sharma-lien:vor:12}).
	For similar reasons, unconditional
	exact (unqualified) path planners in 3D are unknown.

	\ignore{
	Although ``generic'' exact (unqualified) planners exist,
	they are neither practical nor implementable. 
	We need exact (unqualified) planners that are ``explicit'' in the
	sense that its numerical operations are reduced 
	to primitives involving the arithmetic operations.
	Sometimes, root isolation or even solving polynomial systems
	\cite{yap:algebra:bk} are required, but in this case,
	additional effort is needed to pick out the desired roots.
	Often geometric algorithms assume even less elementary
	operations (e.g., ``follow a curve until the first intersection
	with a surface'' in Voronoi diagram algorithms
	\cite{lee-choset:sensor-rod-planning:05}.
	}

	\ignore{
	{\em The algorithms in this paper
	are fully explicit involving numerical approximations
	and which can be guaranteed using interval methods.}
	It would not be ``explicit'' if the algorithm algorithm to say
	``follow a particular 1-dimensional curve'' 
	as in my own old paper.
	}
	\ignore{
	We note that not all exact (unqualified) algorithms are
	equally explicit.
	correct implementations -- in fact it is relatively easy to do,
	using software libraries with ``EGC number types''
	\cite{sharma-yap:crc}.
	Such number types are sophisticated
	forms of big number packages that support exact comparison,
	and can be found in such libraries as
	\cgal \cite{cgal:home}, \leda  \cite{leda:home}
	or \corelib \cite{core:home}. 
	Halperin et al
	\cite{halperin-fogel-wein:bk} describe many such
	exact$_{1,2}$ solutions based on \cgal.
	Two limitations are ``practical efficiency'' and availability
	of exact$_0$ planners.
	}%

	We now address (G1-G3).
	The typical implementation is based on machine arithmetic
	(the IEEE standard), which may satisfy
	(G2) but almost certainly not (G1).
	We regard this as a (G1-G2) trade-off.
	In fact, our implementations here as well as
	in our previous papers
\cite{wang-chiang-yap:motion-planning:15,luo-chiang-lien-yap:link:14,yap-luo-hsu:thicklink:16}
%
	are such machine implementations.  This
	follows the practice in the robotics community, in order to have
	a fair comparison against other implementations.
	Below, we shall expand on our claims about (G1-G3)
	including how to achieve theoretically correct implementation (G1).
	What makes this possible is our replacement of
	``exact (unqualified)'' planners by
	``exact (up to resolution)'' planners, defined below:
	\boxp{
	\dt{Resolution-Exact Path Planning} for robot $R_0$:
	   \\\INPUt: $(\alpha,\beta,\Omega; B_0,\vareps)$
	\\\Indent[8] where $\alpha,\beta\in\cspace(R_0)$ is the start and goal,
		$\Omega\ib\RR^3$ 
	\\\Indent[8] the obstacle set, $B_0\ib\cspace(R_0)$ is a box,
		and $\vareps>0$.
	\\\OUTPUt: Halt with either an $\Omega$-free path from $\alpha$ to
		$\beta$ in $B_0$,
	\\\Indent[8] or \nopath\ satisfying the conditions (P) and (N) below.
	}
	The \dt{resolution-exact planner} (or, $\vareps$-exact planner)
	has an \dt{accuracy constant} $K>1$ (independent of input)
	such that its output satisfies two conditions:
	\bitem
    \item (P) If there is a path (from $\alpha$ to $\beta$ in $B_0$)
	of clearance $K\vareps$, the output {\em must} be\footnote{
	    	For simplicity, we do not require
		the output path to have any particular clearance,
		but we could require clearance $\ge \vareps/K$ as in
		\cite{wang-chiang-yap:motion-planning:15}.
	    }
	 a path.
    \item (N) If there is no path in $B_0$ of clearance $\vareps/K$,
	the output {\em must} be \nopath.
	\eitem
	Here, clearance of a path is the minimum separation of the
	obstacle set $\Omega$ from the robot's footprint on the path.
	Note that the preconditions for (P) and (N) are not exhaustive:
	in case the input fails both preconditions, 
	our planner may either output a path or \nopath.  
	This indeterminacy is essential to escape from exact computation
	(and arguably justified for robotics \cite{sss}). 
	\ignore{%
	    It is a small price to pay for
	making such problems practical, or viewing it positively,
	this indeterminacy matches indeterminacy of physical robots whose
	dimensions, position, sensors and actuators are
	never exact \cite{sss}. 
	}
	The constant $K>1$ is treated in more detail
	in \cite{wang-chiang-yap:motion-planning:15,sss2}.
	But resolution-exactness is just a definition.
	How do we design such algorithms?
	We propose to use subdivision, and
	\ignore{
	    The fundamental operation of subdivision
	algorithms can be viewed as the
	repeated subdivision of the initial box $B_0$ until each box
	$B$ in the subdivision satisfies some box predicate $C(B)$
	(cf.~\cite{cxyz}).
	This process produces a subdivision tree, which is a generalization of
	the familiar quadtrees for boxes in $\RR^2$.
	}
	couple with 
	\dt{soft predicates} to exploit resolution-exactness.
	We replace the classification $C(B)$ by a soft version $\wtC(B)$
	\cite{wang-chiang-yap:motion-planning:15}.
	This leads to a general resolution-exact planner which we call
	\dt{Soft Subdivision Search} (SSS) \cite{sss,sss2}
	that shares many of the favorable properties of sampling
	planners (and unlike exact planners).
	We demonstrated in 
	\cite{wang-chiang-yap:motion-planning:15,luo-chiang-lien-yap:link:14,yap-luo-hsu:thicklink:16}
	that for planar robots with up to 4 DOFs, our
	planners can consistently outperform state-of-the-art 
        sampling planners.  

%
%

	%
	

%
%

	\subsection{What is New: Contributions of This Paper}
%
	In this work, we design 
	$\vareps$-exact planners for rods and rings, with
	accompanying 
	implementation that addresses the desiderata (G0-G3).
	This fulfills a long-time challenge in robotics.
	We are able to do this because of the 
	twin foundations of resolution-exactness and soft-predicates.
	Although we had already used this foundation to implement 
	a variety of planar robots
	\cite{wang-chiang-yap:motion-planning:15,luo-chiang-lien-yap:link:14,yap-luo-hsu:thicklink:16,zhou-chiang-yap:complex-robot:18}
	that can match or surpass state-of-the-art
	sampling methods, it was by no means assured that we can
	extend this success to 3D robots.
	Indeed, the present work required a series of
	technical innovations:
	{\bf (I)}
	One major technical difference from our previous
	work on planar robots
	is that we had to give up the notion of "forbidden orientations"
	(which seems `forbidding' for 3D robots).  We introduced an
	alternative approach based on the ``safe-and-effective''
	approximation of footprint of boxes.
	We then show how to achieve such approximations
	for the rod and ring robots separately.
	{\bf (II)}
	The approximated footprints of boxes are represented by
	what we call $\Sigma_2$-sets
	(Sec.~\ref{sec-4.1-sigma2-set}); this representation supports
	desideratum (G2) for easy implementation. 
	One side benefit of $\Sigma_2$-sets is that they are very
	flexible; thus, we can now easily extend our planners to
	``thick'' versions of the rod or ring.
	In contrast, the forbidden orientation approach
	requires non-trivial analysis to justify
	the ``thick'' version \cite{yap-luo-hsu:thicklink:16}. 
	The trade-off in using $\Sigma_2$-sets is a modest increase in the 
	accuracy constant $K$.
%
	{\bf (III)}
	We also need
	good representations of the 5-DOF configuration
	space.  Here we introduce the square model of $S^2$
	to avoid the singularities in the usual
	spherical polar coordinates
	\cite{lee-choset:sensor-rod-planning:05},
	{\bf and also} to support subdivision in non-Euclidean spaces.
	{\bf (IV)}
	Not only is the geometry in 3D more involved,
	but the increased degree of freedom requires new
	techniques to further improve efficiency. Here, the
	search heuristic based on Voronoi diagrams becomes critical
	to achieve real-time performance (desideratum (G3)).
	\ignore{
	So far, we assumed
	that the planning problem is for a {\em fixed} robot $R_0$.
	But our algorithm and software works for
	the rod robots parametrized by rod length,
	and ring robots parametrized by radius. 
	}

	\ignore{
    \item
	A line segment is hardly a realistic robot.
	But we can thicken the line segment by taking its
	Minkowski sum with a ball
	(see~\cite{yap-luo-hsu:thicklink:16} for a similar procedure for
	2D rods).  The results of this paper easily extend to 
	\dt{thick versions} of our ``thin'' robots, with minimal impact
	on running time.  This again exploits the flexibility of
	resolution-exact planners. 
	Indeed, there are no known exact planners for thick rods.
	}

	\ignore{
	Our 5-DOF spatial robot $R_0$ is pushing current limits for
	subdivision approaches.  Although conventional wisdom says
	that sampling methods can achieve higher DOFs, they are still
	limited to about $5-12$ DOFs according to 
	Choset et al \cite[p.~202]{choset-etal:bk}.
	}%

\noindent {\bf Overview of the Paper} \\
	Section~2 is a brief literature review.
	Section~3 explains an essential preliminary to doing
	subdivision in $S^2$.  Sections~4--6 describe our 
	techniques for computing approximate footprints of rods and rings.
	We discuss efficiency and experimental results
	in Section~7.  We conclude in Section~8. 
	Appendices A-F
%
%
        contain some background and all the proofs.

\sect[lit]{Literature Review}
	\ignore{
	    We would like to understand the limit in terms of
	degrees-of-freedom (DOF) for robots that can be practically handled by
	our methods.  According to Zhang et al
	\cite{zhang-kim-manocha:path-non-existence:08},
	there are no known good implementations of exact motion planners for
	more than 3 DOFs.   On the other hand, the Gamma Group have successfully
	treated 4- and higher-DOF robots using Subdivision Methods.
	Choset et al \cite[p.~202]{choset-etal:bk} suggests that the current
	state of the art in PRM can handle robots of $5-12$ DOFs.
	}
	
	Halperin et al
	\cite{hss:motionplanning:crc:17}
	gave a general survey of path planning.  
	An early survey is \cite{yap:amp:87} where 
	two universal approaches to exact path planning were
	described: cell-decomposition \cite{ss1} and retraction
	\cite{odun-yap:disc:85,odun-sharir-yap:retraction:83,canny:roadmap:93}.
	Since exact path planning is a semi-algebraic problem \cite{ss2},
	it is reducible to general (double-exponential)
	cylindrical algebraic decomposition techniques
	\cite{basu-pollack-roy:bk}.
	But exploiting path planning as a connectivity problem
	yields singly-exponential time (e.g, \cite{eldin-schost:baby-step:11}).
	The case of a planar rod (called ``ladder'') was first studied in
	\cite{ss1} using cell-decomposition. 
	More efficient (quadratic time) methods
	based on the retraction method were introduced in
	\cite{odun-sharir-yap:vorI:86,odun-sharir-yap:vorII:87}.
	On-line versions for a planar rod are also available
	\cite{cox-yap:online-rod:91,choset-mirtich-burdick:sensor-planar-rod:97}.

	Spatial rods were first treated in \cite{ss5}.
	The combinatorial complexity of its free space 
	is $\Omega(n^4)$ in the worst case 
	and this can be closely matched by an
	$O(n^{4+\eps})$ time algorithm \cite{koltun:pianos-flat:05}.
	The most detailed published planner for a 3D rod is
	Lee and Choset
	\cite{lee-choset:sensor-rod-planning:05}.
	They use a retraction approach.
	The paper exposes many useful and interesting details of 
	their computational primitives (see its appendices).
	In particular, they follow a Voronoi edge by a numerical
	path tracking. But like most numerical code, there is no a priori
	guarantee of correctness.
	Though the goal is an exact path planner,
	degeneracies are not fully discussed.
	Their two accompanying videos   
	have no timing or experimental data.

	\ignore{
	Both the sampling approach and our SSS approaches are
	based on what we called ``frameworks'' \cite{sss}, in which
	the overall structure of the planner is reasonably fixed,
	and which requires several subroutines.
	For SSS, it is the design of soft predicates as well as the
	search strategy (encoded as the priority queue's $\getnext()$ method.
	In Zhang et al
	\cite{zhang-kim-manocha:path-non-existence:08}
	they use what they call a ``guided search'' technique
	that goes back to the idea of M-paths of Zhu and Latombe
	\cite{zhu-latombe:hierarchical:91}.
	}
	One of the few papers to address the non-existence
	of paths is Zhang et al \cite{zhang-kim-manocha:path-non-existence:08}.
	Their implementation work is perhaps the closest
	to our current work, using subdivision.
	They noted that 
	``no good implementations are known for general
	robots with higher than three DOFs''.  They
	achieved planners with 3 and 4 DOFs (one of which is a spatial robot).
	Although their planners can detect \nopath,
	they do not guarantee detection (this is impossible without exact
	computation).
	\ignore{
	they adopted the standard notion of resolution-exactness
	\cite[p.~1246]{zhang-kim-manocha:path-non-existence:08}:
	``which
	means they can either find a collision-free path or conclude
	that no such path exists provided that the number of subdivisions is
	sufficiently high or small resolution parameters are chosen.'' 
	Another paper addressing non-existence
	of paths is \cite{basch+3:disconnection:01} which
	\cite{zhang-kim-manocha:path-non-existence:08} says is
	``restricted to very special cases''.
	}

	\ignore{
	``Previously, complete solutions have rarely
	been implemented, mainly due to the lack of the
	nontrivial infrastructure that is needed for such tasks.
	With the recent advancement in the laying out of such
	infrastructure, and in particular with tools now available
	in the software libraries LEDA [MN99] and CGAL [CGAL]
	implementing complete solutions to motion planning
	has become feasible.''
	%
	Despite this optimism, exact implementations in 3D is
	out of the question, simply because there are no explicit
	exact 3D planners. To understand this difficulty, note that
	any exact planner must explicitly handle all 
	degeneracies for the algorithm
	and these are generally unknown in 3D.
	We refer to discussions in \cite{yap-sharma-lien:vor:12}
	where the simpler problem of computing
	the exact Voronoi diagram of a polyhedral set $\Omega$ is open.
	}%

\sect[2]{Subdivision Charts and Atlas for $S^2$}
	{\bf Terminology.}
	We fix some terminology for the rest of the paper.
%
%
	The fundamental \dt{footprint map} $\Fp$ from
	configuration space $\cspace=\cspace(R_0)$ to subsets of $\RR^3$
	was introduced above.
	If $B\ib\cspace$ is any set of configurations, we define
	$\Fp(B)$ as the union of $\Fp(\gamma)$ as $\gamma$
	ranges over $B$.
	Typically, $B$ is a ``box'' of $\cspace$ (see below for 
        its meaning
        in non-Euclidean space $S^2$).
	We may assume $\Omega\ib\RR^3$
	is regular (i.e., equal to the closure of its interior).
	Although $\Omega$ need not be bounded (e.g., it may be the complement
	of a box), we assume its boundary $\partial(\Omega)$
	is a bounded set.  Then $\partial(\Omega)$ is partitioned
	into a set of \dt{(boundary) features}: \dt{corners}
	(points), \dt{edges} (relatively open line segments), or \dt{walls}
	(relatively open triangles). 
	Let $\Phi(\Omega)$ denote the set of features of $\Omega$.
	The (minimal) set of corners and edges is uniquely defined by $\Omega$,
	but walls depend on a triangulation of $\partial\Omega$.
	If $A,B\ib\RR^3$, define their \dt{separation}
	$\Sep(A,B)\as \inf\set{\|a-b\|: a\in A, b\in B}$ where $\|a\|$ is
	the Euclidean norm.
	The \dt{clearance} of $\gamma$ is $\Sep(\Fp(\gamma),\Omega)$.
	Say $\gamma$ is \dt{$\Omega$-free} (or simply \dt{free})
	if it has positive clearance.
	Let $\cfree=\cfree(\Omega)$ be the set of $\Omega$-free
	configurations.
	The \dt{clearance} of a path $\mu:[0,1]\to\cspace$
	is the minimum clearance
	attained by $\mu(t)$ as $t$ ranges over $[0,1]$.

        {\bf Subdivision in Non-Euclidean Spaces.}
	Our $\cspace$ has an Euclidean part 
	($\RR^3$) and a non-Euclidean part ($S^2$).
	We know how to do subdivision in $\RR^3$ but it
	is less clear for $S^2$.
	Non-Euclidean spaces can be represented either 
	(1) as a submanifold of $\RR^m$ for some $m$
	(e.g., $SO(3)\ib\RR^9$ viewed as orthogonal matrices)
	or (2) as a subset of $\RR^m$
	subject to identification (in the sense of quotient topology
	\cite{munkres:topology:bk}).  
	A common representation of $S^2$
	(e.g., \cite{lee-choset:sensor-rod-planning:05})
	uses a pair of angles (i.e., spherical polar coordinates)
	$(\theta,\phi)\in [0,2\pi]\times [-\pi/2,\pi/2]$
	with the identification
	$(\theta,\phi)\equiv(\theta',\phi')$
	iff $\set{\theta,\theta'}=\set{0,2\pi}$
		or $\phi=\phi'=\pi/2$ (North Pole)
		or $\phi=\phi'=-\pi/2$ (South Pole).
	Thus an entire circle of values $\theta$ 
	is identified with each pole, causing severe distortions
	near the poles which are singularities.  So
	the numerical primitives in  
	\cite[Appendix F]{lee-choset:sensor-rod-planning:05}
	have severe numerical instabilities.

	To obtain a representation of $S^2$ without singularities,
	we use the map \cite{sss2}
		\[q\in\RR^3\mapsto \whq \as q/\|q\|_{\infty}\]
	whose range is the boundary of a 3D cube
	$\sqs\as \partial([-1,1]^3)$.
	This map is a bijection when its domain is restricted to $S^2$,
	with inverse map $q\in\sqs\mapsto \ol{q}\as q/\|q\|_2\in S^2$.
	Thus $\ol{\whq}$ is the identity for $q\in S^2$.
	We call $\sqs$ the \dt{square model} of $S^2$.
	%
	We view $S^2$ and $\sqs$ as metric spaces: 
	$S^2$ has a natural metric whose geodesics are
	arcs of great circles.  The geodesics on $S^2$
	are mapped to the corresponding polygonal geodesic paths on $\sqs$ by 
      %
      %
        $q\mapsto \whq$.  Define the constant
		\[C_0\as \sup_{p\neq q\in S^2} \set{
		\max\set{
		    	\frac{d_2(p,q)}{\wh{d}_2(\wh{p},\wh{q})},\;
			\frac{\wh{d}_2(\wh{p},\wh{q})}{d_2(p,q)} }}
		\]
	where $d_2$ and $\wh{d}_2$ are the metrics on $S^2$ and $\sqs$
	respectively.  Clearly $C_0\ge 1$.
	Intuitively, $C_0$ is the largest
	distortion factor produced by the map $q\mapsto \whq$
	(by definition the inverse map has the same factor).
	\blem
	$C_0=\sqrt{3}$.
	\elem
\remove{
	\bpf Let $B$ be the ball whose boundary is $S^2$ and $C=[-1,1]^3$.
	Then $B/\sqrt{3}\ib C \ib B$.   From any geodesic $\alpha$ of $S^2$,
	we obtain a corresponding geodesic $\alpha'$ on the surface of
	$B/\sqrt{3}$, and a geodesic $\wh{\alpha}$ of $\sqs=\partial(C)$. 
	Observe
	that $|\alpha'| \le |\wh{\alpha}| \le |\alpha|$ where $|\cdot|$
	is the length of a geodesic.  But $|\alpha| =\sqrt{3}|\alpha'|$.
	This proves our bound.  This bound is tight because for geodesic
	arcs in arbitrary small neighborhoods of the centers of the 6 faces
	of $\sqs$, the bound is arbitrarily close to $\sqrt{3}$.
	\epf
}
	The proof in Appendix~\ref{appendixC-square-model} 
	also shows that the worst distortion
	is near the corners of $\sqs$.
	The constant $C_0$ is 
        one of the 4 constants
%
%
	that go into the ultimate accuracy
	constant $K$ in the definition of $\vareps$-exactness
	(see \cite{sss2} for details).
	
	It is obvious how to do subdivision in $\sqs$.
	This is illustrated in \refFig{cone-disc-ring}(b).
	After the first subdivision of $\sqs$ into 6 faces,
	subsequent subdivision is just the usual quadtree subdivision
	of each face.  We interpret the subdivision of $\sqs$
	as a corresponding subdivision of $S^2$.  In
	\cite{sss2}, we give the general framework using the notion of
	\dt{subdivision charts and atlases}
	(borrowing terms from manifold theory).

	\ignore{
	In order to do subdivision in $S^2$, we introduce the
	\dt{subdivision charts and atlases}
	(borrowing terms from manifold theory).  
	First, we subdivide $\sqs$ into its 6 faces, denoted
	$S_i$ ($i\in I$) where
	$I \as \set{+,-}\times \set{x,y,z}$ serves as the
	index set for the 6 faces.  Each $i\in I$ corresponds
	to a semi-axis.  
	This is illustrated in \refFig{cone-disc-ring}(b).
	Intuitive, each $S_i$ is the root of a
	subdivision tree (actually quadtree).
	A box $B$ in one of these subdivision trees
	represents a corresponding region of $S^2$.  More precisely,
	we define six maps called \dt{charts},
	$\mu_i:[-1,1]\to S^2$ such that the range of these six charts
	covers $S^2$.   A more explicit description is to write
	$\mu_i(a,b)\as \wh{m_i(a,b)}$ where each $m_i: [-1,1]\to S_i$
	in a natural way.
	E.g., if $i=+x$ then $m_i(a,b)=(1,a,b)$,
	and $i=-y$ then $m_i(a,b)=(a,-1,b)$, etc.
	We can now do subdivision for $S^2$ via these charts
	(the initial subdivision of $S^2$ simply splits into the
	6 $S_i$'s, but subsequent splits creates four congruent
	subsquares, as in a quadtree).
	The set $\set{\mu_i: i\in I}$ charts is called
	an \dt{atlas} for $S^2$;
	see \cite{sss2} for a general treatment of 
	subdivision charts and atlases.
	}

	\ignore{ 
	Fix some polyhedral set $\Omega\ib\RR^3$ of obstacles.
	We assume that the $\Omega$ is ``regular'' in the sense
	that it is equal to the closure of its interior.
	The \dt{separation} between two sets $S,T\ib\RR^3$ is
		$\Sep(S,T)=\inf\set{\|s-t\|: s\in S, t\in T}$.
	Note that $\|\cdot\|=\|\cdot\|_2$ is the Euclidean norm.
	When $S$ is fixed, we also write $\Sep_S:\RR^3\to\RR_{\ge 0}$
	for the function $\Sep_S(q)\as \Sep(S,\set{q})$.  
	The \dt{$\Omega$-clearance}
	function 
	$\Cl_\Omega: \cspace \to \RR_{\ge 0}$   
	is given by
	$\Cl_\Omega(\gamma) \as \Sep(AB[\gamma],\Omega)$.
	We omit the subscript $\Omega$ when it is understood.
	We say $\gamma$ is \dt{free} if $\Cl(\gamma)>0$.
	}%
	
	\ignore{
	{\bf Boxes of $\RR^3\times S^2$.}
	By ``box'' of $\RR^3\times S^2$, we mean a set
	$B\ib \cspacehat =\RR^3\times \sqs$ of the form $B=B^t\times B^r$
	where $B^t$ is the usual axis-parallel
	box in $\RR^3$, and $B^r$ is
	either equal to $\sqs$ or equal to a box of some face of $S_i$
	($i\in I$).   In the latter case, we represent $B^r$
	as a box in $[-1,1]$ together with an indicator $i\in I$.
	We call $B^t$ and $B^r$ the 
	\dt{translational} and \dt{rotational} components of $B$.
	Two $d$-dimensional boxes are \dt{adjacent} if
	their intersection is a non-degenerate $d-1$-dimensional set.
	Besides the adjacencies of boxes of $S_i$ that comes from subdivision,
	each box on the boundary of $S_i$ is also adjacent with some $S_j$
	($i\neq j$) which are relatively easy to maintain.
	%
	Let $m_{B}$ and $r_{B}$ denote the center (``midpoint'') and
	radius of $B^t$ (radius is the distance from the center to any corner
	of the cube).   In other words, $m_B$ and $r_B$
	ignores the rotational part.  There are 2 kinds of splits:
	\dt{T-split($B$)}
	splits $B^t$ into 8 congruent subboxes,
	which are paired with the unsplit $B^r$ to produce $8$ children.
	\dt{R-split($B$)}
	has 2 cases: if $B^r=\sqs$, then it splits $\sqs$ into $6$ faces.
	Otherwise, $B^r$ is a square of $\sqs$, and we split it into
	$4$ congruent squares.  In either case, each of the squares
	are then paired with $B^t$ to form the children of $B$.
	}%

\sect[3]{Approximate Footprints for Boxes in $\RR^3\times S^2$}
\label{se-let-us-design}
	We focus on soft predicates because, in principle,
	once we have designed and implemented such a predicate,
	we already have a rigorous and complete planner within the
	\dt{Soft Subdivision Search} (SSS) framework
	\cite{wang-chiang-yap:motion-planning:15,sss2}.
	For convenience, the SSS framework is summarized in Appendix~\ref{app-SSS}. 
	As noted in the introduction, our soft predicate $\wtC$ classifies
	any input box $B\ib\cspace$ into one 3 possible values.
	A key idea of our 2-link robot work
	\cite{luo-chiang-lien-yap:link:14,yap-luo-hsu:thicklink:16}
	is the notion of ``forbidden orientations''
	(of a box $B$, in the presence of $\Omega$).
	The same concept may be attempted
	for $\RR^3\times S^2$, except that the details
	seem to be formidable to analyze and to implement.
	Instead, this paper introduces a direct approximation of the 
	\dt{footprint of a box},
	$\Fp(B)\as\bigcup \set{\Fp(\gamma):\gamma\in B}$.
	We now introduce
	$\wtFp(B)\ib\RR^3$ as the \dt{approximate footprint},
	and discuss its properties.
	This section is abstract, in order to expose the
	mathematical structure of what is needed to achieve
	resolution-exactness for our planners.  The reader might
	peek at the next two sections to see the instantiations
	of these concepts for the rod/ring robot.

	To understand what is needed of this approximation,
	recall that our approach to soft predicates is based
	on the ``method of features''
	\cite{wang-chiang-yap:motion-planning:15}.
	The idea is to maintain a set $\wtphi(B)$
	of approximate features for each box $B$.
	We softly classify $B$ as $\wtC(B)=\mixed$
	as long as $\wtphi(B)$ is non-empty; otherwise,
	we can decide whether $\wtC(B)=C(B)$ is \free\ or \stuck.
        This decision is relatively easy in 2D, but is more 
	involved in 3D and detailed in
	Appendix~\ref{appendixC-classify-box}. 
   	For correctness of this procedure, we require
                \beql{incl2}
		\wtphi(B/\sigma) \ib \phi(B) \ib \wtphi(B).
		\eeql
	Here $\sigma>1$ is some global constant and
	``$B/\sigma$'' denotes the box $B$ shrunk by factor $\sigma$.
	Basically, \refeq{incl2} guarantees that our soft predicate
	$\wtC(B)$ is conservative and $\sigma$-effective (i.e.,
	if $B$ is free then $\wtC(B/\sigma)=\free$).
	For computational efficiency, we want the
	approximate feature sets to have \dt{inheritance property}, i.e.,
	\beql{incl3}
		\wtphi(B)\ib \wtphi(parent(B)).
	\eeql

\remove{ 
%
%
Now we discuss how to classify a box $B$ as \free\ or
\stuck\ when its feature set is empty.
Suppose $\Omega$ is given as the union of a set of polyhedra
%
%
that may overlap (this situation arises in Sec.~\ref{se-expt}). 
Let $B'$ be the parent of $B$, then the feature set $\wtphi(B')$ is
non-empty. For each obstacle polyhedron $P$ in $\wtphi(B')$, we find
the feature $f \subseteq \partial P$ closest to $m_B$ and use $f$ to
decide whether $m_B$ is outside $P$. Then $m_B$ is outside $\Omega$
(and $B$ is \free) iff $m_B$ is outside all such polyhedra $P$. 

\remove{ 
\noindent {\bf ***??? YJC: Note: $m_B$ must be outside *all* such
  polyhedra $P$. So for *each* $P$ we must find the closest feature
  $f$ of $P$ to $m_B$, rather than just the *one* closest feature $f$
  of $\Omega$.} \\
} 

To find the feature $f \subseteq \partial P$ closest to $m_B$, we
first find among the corners of $P$ the one $f_c$ that is the closest.
Then among the edges of $P$ incident on $f_c$, we check if there exist
edges $e$ that are even closer (i.e., $\Sep(e, m_B) < ||f_c - m_B||$;
$\Sep(e, m_B)$ is defined by a point interior to $e$) and if so pick
the closest one $f_e$. Finally, if $f_e$ exists, we repeat the process
for faces of $P$ incident on $f_e$ and pick the closest one $f_w$ (if
it exists). The closest feature $f$ is set to $f_c$ then updated to
$f_e$ and to $f_w$ accordingly if $f_e$ (resp.\ $f_w$) exists.

Given the feature $f \subseteq \partial P$ closest to $m_B$, we can
easily determine if $m_B$ is interior or exterior of $P$ when $f$ is a
wall or an edge.  When $f$ is a corner, it is slightly more
involved.  
%
%
We will classify a corner $f$ to be \dt{pseudo-convex} (resp.,
\dt{pseudo-concave}) if there exists a closed half space $H$ such that
(1) $f \in \partial H$, (2) $H$ contains an edge of $P$ incident on
$f$, and (3) for any small enough ball $\Delta$ centered at $f$, we
have that $(\Delta \cap P)\setminus H$ is disjoint from $P$ (resp.,
contained in $P$).  
%
%
Note that if $f$ is locally convex (resp., locally
concave) then it is pseudo-convex (resp., pseudo-concave).
We call a corner $f$ an \dt{essential corner} if for all balls
$\Delta$ centered at $f$, $\Delta \cap \partial P$ is not a planar
set.  We may assume that our corners are essential; as consequence, no
corner can be both pseudo-convex and pseudo-concave.  However, it
is possible that a corner is neither pseudo-convex nor pseudo-concave;
we call such corners \dt{mixed}.
%
%
The lemma below enables us to avoid the difficulty of mixed corners.
\blem
Let $q\notin\partial P$ and $C$ a corner of $P$.
If $C$ is the point in $\partial P$ closest to $q$,
i.e., $\Sep_{\partial P}(q)=\|q-C\|$, then $C$ is either
pseudo-convex or pseudo-concave.  Hence $C$ cannot be a mixed corner.
Moreover, $q \in P$ iff $C$ is pseudo-concave.
\elem
\bpf
%
Let $\Delta$ be the ball centered at $q$ with radius $\|q-C\|$.
Since $\Sep_{\partial P}(q)=\|q-C\|$, we have $\Delta \cap \partial P =\set{C}$.
Let $H$ be the closed half-space such that $\partial H$
is tangential to $\Delta$ at the point $C$, and $q\notin H$.
This $H$ is a witness to either the 
pseudo-convexity or pseudo-concavity of $C$.  
In particular, $C$ is pseudo-concave iff $q\in P$.
\epf

\remove{
Note: Using the approach (def. of pseudo-convexity/pseudo-concavity & the Lemma & Pf above),
      we do NOT need to worry about mixed corners, i.e.,
      In Algorithm 4 (in the appendix of scg2017 write-up (abs.tex there)), 
      when the closest feature is a corner, it will NEVER be a mixed corner.

This Algorithm 4 is NO Longer Needed to be presented in this paper!

For fixing the code, Algorithm 4 needs the following changes for this part.

Changes Needed:

1. Change the conditions of convex/concave corner to
   pseudo-convex/pseudo-concave corner.

2. Remove the case of mixed corner.

3. $f$ being pseudo-convex/pseudo-concave does not give 
   \free/\stuck\ to the box $B$ yet; it only decides that 
   $m_B$ is outside/inside the current polyhedron $P$. 

** Here is the portion of Algorithm 4 (copied here), the case of closest feature being a corner:
%
\noindent         
        {\tt //} This is the case where $f$ is a corner \\
%
           {\bf if} $f$ is a pseudo-convex corner {\bf return} $\free$ \\  
%
           {\bf elseif} $f$ is a pseudo-concave corner {\bf return} $\stuck$ \\ 
%
\remove{
           {\bf else} {\tt //} $f$ is a mixed corner \\
%
%
            Find a wall $w'$ with equation $L'$ incident to $f$
            and has the minimum angle with plane $\Pi: \overrightarrow{m(B)-f} \cdot (x - f) = 0 $\\
%
            {\bf return} the sign of $L'(m(B))$ \\
} 
%
%
} 
%
%
} 

	\ignore{%
	    USE THIS TEXT:
	The idea is that if $B'$ is the parent of $B$,
	then $\wtphi(B')$ is non-empty when $\wtphi(B)$ is empty.
	Let $f$ be the closest feature in $\wtphi(B')$ to $m_B$:
	if $f$ is a wall or an edge, we can easily determine if $m_B$
	is interior or exterior of the obstacle sets, and thus
	$B$ is \stuck\ or \free\ respectively.  If $f$ is a corner
	then again we can make a decision if $f$ is a convex or
	concave corner.  Unfortunately, in 3D, $f$ may not be
	convex or concave. The following lemma solves the difficulty:

	COPY LEMMA from the old writeup section "Soft Pred and the Method of
	Features"...
	}%
%
%

	We now show what this computational scheme demands of our
	approximate footprint.
	Define the \dt{exact feature set} of box $B$ as usual:
	$\phi(B)\as\set{f\in\Phi(\Omega): f\cap \Fp(B)\neq\es}$
	and (tentatively) the \dt{approximate feature set} of box $B$ as
	\beql{def-wtphi} 
	\wtphi(B)\as\set{f\in\Phi(\Omega): f\cap \wtFp(B)\neq\es}.
	\eeql
	The important point is that $\wtFp(B)$ is defined
	prior to $\wtphi(B)$.
	We need the fundamental inclusions
		\beql{incl1}
		\wtFp(B/\sigma) \ib \Fp(B) \ib \wtFp(B).
		\eeql
	Note that this immediately implies \refeq{incl2}.
Unfortunately, \refeq{def-wtphi} and \refeq{incl1}
together do not guarantee inheritance, i.e., \refeq{incl3}.
Instead, we define $\wtphi'(B)$ recursively as follows:
	\beql{wtphi'}
	\wtphi'(B) \as \clauses{
	    	\set{f\in\Phi(\Omega): f\cap \wtFp(B)\neq\es}
			& \textrm{if $B$ is the root,}\\
		\set{f \in \wtphi'(parent(B)): f\cap \wtFp(B)\neq\es}
			& \textrm{else.}}
	\eeql
Notice that this only defines $\wtphi'(B)$ when $B$ is an aligned
box (i.e., obtained by recursive subdivision of the root box).
But $B/\sigma$ is never aligned when $B$ is aligned,
and thus $\wtphi'(B/\sigma)$ is not captured by \refeq{wtphi'}.
Therefore we introduce a parallel definition:
	\beql{wtphi''}
	\wtphi'(B/\sigma) \as \clauses{
	    	\set{f\in\Phi(\Omega): f\cap \wtFp(B/\sigma)\neq\es}
			& \textrm{if $B$ is the root,}\\
		\set{f \in \wtphi'(parent(B)/\sigma): f\cap \wtFp(B/\sigma)\neq\es}
			& \textrm{else.}}
	\eeql

Now, $\wtphi'(B)$ satisfies \refeq{incl3}.
But does it satisfy \refeq{incl2}, which is necessary for correctness?
This is answered affirmatively by the following lemma
(see proof in Appendix~\ref{app-pf-lemma2}):  
\blem \label{lem-inheritance-feature-set-auto-inclusions}
If the approximate footprint $\wtFp(B)$ satisfies Eq.~\refeq{incl1},
%
%
%
then $\wtphi'(B)$ satisfies Eq.~\refeq{incl2}, i.e.,
	\[\wtphi'(B/\sigma) \ib \phi(B) \ib \wtphi'(B).\]
\elem  
Since $\wtphi'(B)$ has all the properties we need,
we have no further use for the definition of $\wtphi(B)$
given in \refeq{def-wtphi}.
Henceforth, we simply write ``$\wtphi(B)$''
to refer to the set $\wtphi'(B)$ defined in~\refeq{wtphi'} and~\refeq{wtphi''}.
%

\remove{ 
\bpf
Let $\wtFp(B)$ be the approximate footprint defining $\wtphi(B)$ in
$\wtphi(B)\as\set{f\in\Phi(\Omega): f\cap \wtFp(B)\neq\es}$.  We will
show that $\wtFp(B)$ satisfies Eq.~\refeq{incl1}, i.e.,
$\wtFp(B/\sigma) \ib \Fp(B) \ib \wtFp(B)$, which equivalently shows
that $\wtphi(B)$ satisfies Eq.~\refeq{incl2} (i.e., $\wtphi(B/\sigma)
\ib \phi(B) \ib \wtphi(B)$).
From the construction of $\wtphi(B)$, we have $\wtFp(B) = \wtFp'(B)$
when $B$ is the root, and $\wtFp(B) = \wtFp'(B) \cap \wtFp(parent(B))$
otherwise.
The case when $B$ is the root is trivial, since $\wtFp'(B)$ satisfies
Eq.~\refeq{incl1} and thus $\wtFp(B) = \wtFp'(B)$ satisfies
Eq.~\refeq{incl1} as well.
Now suppose $B$ is not the root (and thus $\wtFp(B) = \wtFp'(B) \cap
\wtFp(parent(B))$). \\
(1) First we prove that $\Fp(B) \ib \wtFp(B)$.  By the condition of
the theorem, $\wtFp'(B)$ is a superset of $\Fp(B)$.  It suffices to
show that $\wtFp(parent(B))$ is a superset of $\Fp(B)$. But
$\wtFp(parent(B))$ is a superset of $\Fp(parent(B))$ (initially at the
root and inductively top down), which in term is a superset of
$\Fp(B)$. \\
(2) Finally we prove that $\wtFp(B/\sigma) \ib \Fp(B)$.  By the
condition of the theorem, $\wtFp'(B/\sigma) \ib \Fp(B)$, and
$\wtFp(B/\sigma) = \wtFp'(B/\sigma) \cap \wtFp(parent(B/\sigma))$ is
an even smaller set than $\wtFp'(B/\sigma)$.
\epf
} 
%

%
%
%
        {\bf Geometric Notations.}
	We will be using planar concepts like circles, squares, etc,
	for sets that lie in some plane of $\RR^3$.  We shall
	call them \dt{embedded} circles, squares, etc.
	By definition, if $X$ is an embedded object then
	it defines a unique plane $Plane(X)$ (unless $X$ lies in a line).
	Let $Ball(r,c)\ib\RR^3$ denote a ball of radius $r$
	centered at $c$.  If $c$ is the origin, we simply write $Ball(r)$.
	Suppose $X\ib \RR^3$ is any non-empty set.
	Let $Ball(X)$ denote the \dt{circumscribing ball} of $X$, defined
	as the smallest ball containing $X$.
	Next, if $c\notin X$ then $Cone(c, X)$ denotes 
	the union of all the rays from $c$ through points in $X$,
	called the \dt{cone} of $X$ with \dt{apex} $c$.
	We consider two cases of $X$ in this cone definition:
	if $X$ is a ball, then $Cone(c,X)$ is called
	a \dt{round cone}.
	If the radius of ball $X$ is $r$ and the distance from the center
	of $Ball(X)$ to $c$ is $h\ge r$, then call $\arcsin(r/h)$
	the \dt{half-angle} of the cone; note that the angle at the
	the apex is twice this half-angle.
	If $X$ is an embedded square, we call $Cone(c,X)$
	a \dt{square cone}, and the ray from $c$ through the center
	of the square is called the \dt{axis} of the square cone.
	If $P$ is any plane that intersects the axis of a square cone
	$Cone(c,X)$, then $P\cap Cone(c,X)$ is a square iff $P$ is parallel to
	square $X$.
	A \dt{ring} (resp., \dt{cylinder})
	is the Minkowski sum of an embedded circle (resp., a line) with a ball. 
	Finally consider a box $B=B^t\times B^r\ib \RR^3\times \sqs$
	where 
        $B^t$ and $B^r$ are the \dt{translational} and \dt{rotational}
        components of $B$, and
        $B^r$ is 
	either $\sqs$ or a subsquare of a face of $\sqs$.
        We let $m_{B}$ and $r_{B}$ denote the center and
	radius (distance from the center to any corner) of $B^t$.
	The \dt{cone of $B$}, denoted $Cone(B)$, is the round
	cone $Cone(m_B, Ball(m_B+B^r))$.
	If the center of square $m_B+B^r$ is $c$ and width of $B^r$ is $w$,
	then $Cone(B)$ is just $Cone(m_B, Ball(c, w/\sqrt{2}))$.

\ssect{On $\Sigma_2$-Sets} \label{sec-4.1-sigma2-set}
	Besides the above inclusion properties of $\wtFp(B)$,
	we also need to decide
	if $\wtFp(B)$ intersects a given feature $f$.
	We say $\wtFp(B)$ is ``nice'' if there are
	intersection algorithms that are
	easy to implement (desideratum G2) and practically efficient 
	(desideratum G3).
	We now formalize and generalize some ``niceness''
	properties of $\wtFp(B)$ that were implicit in our previous work
	(\cite{wang-chiang-yap:motion-planning:15,luo-chiang-lien-yap:link:14,yap-luo-hsu:thicklink:16},
	especially \cite{zhou-chiang-yap:complex-robot:18}).

	An \dt{elementary set} (in $\RR^3$)
	is defined to be one of the following sets or their complements:
	half space, ball, ring, cone or cylinder.
	Let $\Elem$ (or $\Elem_3$) denote the set of elementary sets in
        $\RR^3$.
	In $\RR^2$, we have a similar notion of elementary sets
	$\Elem_2$ comprising half-planes, discs or their complements.
	All these elementary sets are defined by a
	single polynomial inequality -- so technically, they are
	all ``algebraic half-spaces''.
	The sets in $\Elem$ are evidently ``nice'' 
	(niceness of a ring has some
	subtleties -- see Sec.~\ref{sect:ring}).
We next extend our collection of nice sets:
define a \dt{$\Pi_1$-set} to be a finite intersection of
elementary sets.  We regard a $\Pi_1$-set $S=\cap_{i=1}^n S_i$
to be ``nice'' because we can easily 
check if a feature $f$ intersects $S$ 
%
%
%
%
%
%
    by a simple while-loop (see below).
%
%
Notice that $\Pi_1$ contains all convex polytopes in $\RR^3$.
Our definitions of $\wtFp(B)$ in
\cite{wang-chiang-yap:motion-planning:15,luo-chiang-lien-yap:link:14,yap-luo-hsu:thicklink:16}
are all $\Pi_1$-sets.  But in 
\cite{zhou-chiang-yap:complex-robot:18},
we make a further extension:
define a \dt{$\Sigma_2$-set} to be a finite union of the $\Pi_1$-sets,
i.e., each $\Sigma_2$-set $S$ has the form
		\beql{sigma2}
		S = \bigcup_{i=1}^n \bigcap_{j=1}^{m_i} S_{ij}
		\eeql
where $S_{ij}$'s are elementary sets.  
We still say such an $S$ is ``nice'' since checking if
a feature $f$ intersects $S$ can be written
in a doubly-nested loop 
%
%
   (see below).
Although this intersection is more expensive to check than
with a $\Pi_1$-set, it may result in fewer subdivisions and better
efficiency in the overall algorithm.  Thus, there is an accuracy-efficiency
trade-off.
	%
	{\em Good approximations of footprints are harder to do accurately in
	3D, and the extra power of $\Sigma_2$ seems critical.}

	We can put all these in the framework of
	a well-known\footnote{ 
		From mathematical analysis,
		constructive set theory and complexity theory.
	} construction of an
	infinite hierarchy of sets, starting from some initial
	collection of sets.
	If $\Delta$ is any collection of sets,
	let $\Pi(\Delta)$ denote the collection of finite intersections of
	sets in $\Delta$; similarly,
	$\Sigma(\Delta)$ denotes the collection of finite unions of
	sets in $\Delta$.
	Then, starting with any collection $\Delta_1$
	of sets, define the infinite hierarchy of sets:
		\beql{hierarchy}
		\qquad\qquad \Sigma_i, \Pi_i, \Delta_i\qquad (i\ge 1)
		\eeql
	where $\Sigma_i \as\Sigma(\Delta_{i})$,
	$\Pi_i \as\Pi(\Delta_{i})$, and
	$\Delta_{i+1} \as\Sigma_i\cup \Pi_i$.
%
%
	An element of $\Sigma_i$ or $\Pi_i$
	is simply called a $\Sigma_i$-set or a $\Pi_i$-set.
	
	\ignore{
	In our previous work, $\Delta_0$ was the collection of
	elementary sets in $\RR^2$, and predicates are defined
	with the help of ``swept areas'' which turns out to
	be $\Pi_1$-sets.
	In this paper, we take $\Delta_0$
	to be $\Elem$ and we shall be interested mainly in $\Sigma_2$.
	Clearly, a $\Sigma_2$-set is a finite union of a finite
	intersection of elementary sets.
	However, there is a complexity-accuracy
	trade-off between using approximate $\Pi_1$-set and exact
	$\Sigma_2$-set.  This trade-off should be
	studied empirically as the ``practically efficient''
	criteria of (G3).  
	Note that a $\Sigma_2$-set $S$ has a description of the form
		\beql{sigma2}
		S = \bigcup_{i=1}^n \bigcap_{j=1}^{m_i} S_{ij}
		\eeql
	for some $n$ and $m_i$'s, and $S_{i,j}$ are elementary sets.
	}%
	We call \refeq{sigma2} a \dt{$\Sigma_2$-decomposition} of $S$,
        where $\Delta_1 \as \Elem$.
%
%
	Note that this decomposition may not be unique, but in the
	cases arising from our simple robots,
	there is often an obvious optimal description.
	Moreover, $n$ and $m_i$'s are small constants.
	We can construct new sets by manipulating such a decomposition,
        e.g., replacing each $S_{ij}$ by its \dt{$\tau$-expansion},
%
%
%
%
        i.e., $S_{ij} \oplus Ball(\tau)$ (where $\oplus$ denotes the Minkowski sum),
        which remains elementary.  
%
%
	Under certain conditions, the corresponding
	set is a reasonable approximation to 
%
%
%
        $S\oplus Ball(\tau)$.
	If so, we can generalize the corresponding
	soft predicate to robots with thickness $\tau$.

	Once we have a 
%
%
%
        $\Sigma_2$-decomposition
        of $\wtFp(B)$,
	we can implement the intersection test with
	relative ease (G2) and quite efficiently (G3).
	For instance we can test intersection of the
	set $S$ in \refeq{sigma2} with a feature $f$ by
	writing a doubly nested loop.  At the beginning of the inner loop,
	we can initialize a set $f_0$ to $f$.
	Then the inner loop amounts to the update
			``$f_0 \ass f_0\cap S_{ij}$''
	for $j=1\dd m_i$.  If ever $f_0$ becomes empty, we know
	that the set $S_i=\bigcap_{j=1}^{m_i} S_{ij}$
%
%
	has empty intersection with $f$.
	The possibility of such representations is by no means
	automatic but in the next two sections we verify that they
	can be achieved for our rod and ring robots.  These sections
	make our planners fully ``explicit'' for an implementation.

\sect[4]{Soft Predicates for a Rod Robot}
	In this section, $R_0$ is a rod with length $r_0$;
	we choose one endpoint of the rod as the rotation center.
    Let $B=B^t\times B^r\ib\RR^3\times
	\sqs$ be a box.  Our main goal is to define
	approximate footprint $\wtFp(B)$,
	and to prove the basic
	inclusions in Eqs.~\refeq{incl1} and~\refeq{incl2}. 
	This turns out to be a $\Pi_1$-set (we also indicate a
	more accurate $\Sigma_2$-set.)

	
	    	\begin{figure}[htb]
	    	  \begin{center}
		   \scalebox{0.20}{
	    	     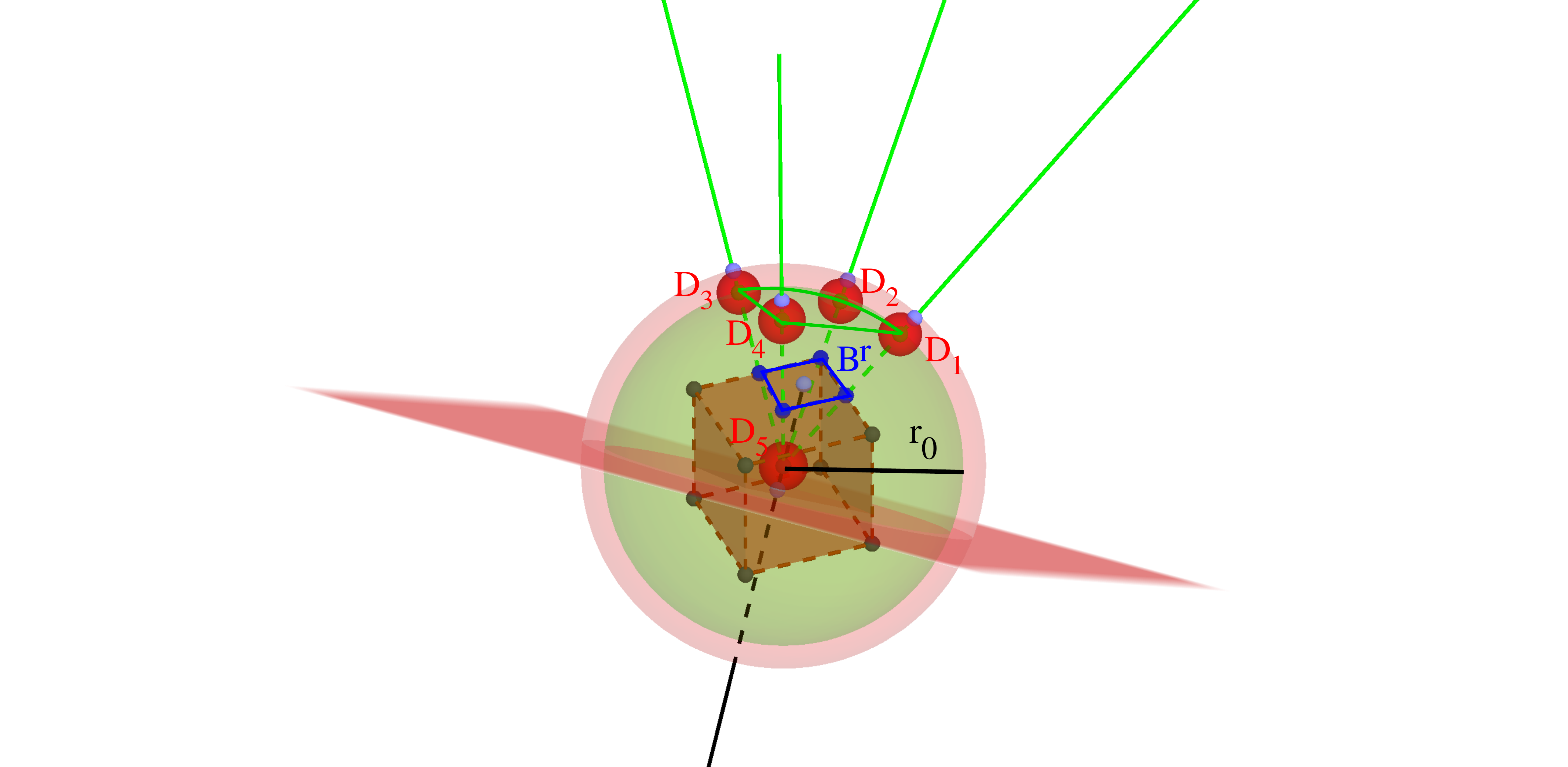}
	    	   \caption{Rod Robot:
		$\Fp(B) = \Fp_{\,0}(B)\oplus (B^t-m_B)$ where
		$\Fp_{\,0}(B)$ is indicated by the four green rays.
%
}
	    	   \label{fig:rod-Minkowski-sum}
	    	  \end{center}
	    	\end{figure} 	 
	It is useful to define
	the \dt{inner footprint} of $B$, 
		$\Fp_{\,0}(B)$, as $\Fp(m_B\times B^r)$. 
%
%
%

	This set is the intersection of a ball and a square cone:
		\beql{sqcone}
			 \Fp_{\,0}(B) = Ball(r_0,m_B)\cap Cone(m_B, B^r+m_B).
		\eeql
	The edges of this square cone
	is shown as green lines in \refFig{rod-Minkowski-sum};
	furthermore, the brown box is $\sqs+m_B$
	(translation of $\sqs$ so that it is centered at $m_B$).
	Note that the box footprint $\Fp(B)$
	is the Minkowski sum of $\Fp_{\,0}(B)$ with $B^t-m_B$ (the
	translation of $B^t$ to make it centered at the origin).
	It is immediate that 
			\[\Fp_{\,0}(B) \ib Cone(B).\]
	Thus we may write
		$Cone(m_B, B^r+m_B)$ as the intersection
	of four half spaces $H_i$ ($i=1\dd 4$).
	Let
		$Cone^{(+r_B)}(m_B, B^r+m_B)$ 
	denote the intersection of the expanded
	half-spaces, $H_i\oplus Ball(r_B)$ ($i=1\dd 4$).
	In general,
		$Cone^{(+r_B)}(m_B, B^r+m_B)$ 
	is not a cone (it may not have a unique ``apex'').
	Similarly we ``expand'' the inner footprint of \refeq{sqcone}
	into 
		\beql{sqcone2}
		``\wtFp(B)\textrm{''} \as
		Ball(r_0+r_B,m_B)\cap Cone^{(+r_B)}(m_B, B^r+m_B).
		\eeql
	We use quotes for ``$\wtFp(B)$'' in \refeq{sqcone2} because
	we view it as a candidate for an approximate
	footprint of $B$.  Certainly, it has the desired property
	of containing the exact footprint $\Fp(B)$.  Unfortunately,
	this is not good enough.  To see this,
	let $\theta$ be the half-angle of the round cone
	$Cone(B) = Cone(m_B,Ball(B^r+m_B))$.
	Then Hausdorff distance of
	``$\wtFp(B)$'' from $\Fp(B)$ can be arbitrarily
	big as $\theta$ becomes arbitrarily small.
	Indeed $\theta$ can be arbitrarily small because it
	can be proportional to the input resolution $\vareps$.
	We conclude that such a planner is not resolution-exact.
	To fix this problem, we finally define
\beql{sqcone3}
		\wtFp(B) \as ``\wtFp(B)\textrm{''} \cap H_0
\eeql
	where $H_0$ is another half space.  A natural choice for $H_0$
	is the half-space ``above'' the pink-color plane of
	\refFig{rod-Minkowski-sum}, defined as the plane
	normal to the axis of cone $Cone(B)$ and at distance $r_B$
	``below'' $m_B$.  
%
%
We can also use 
        the ``horizontal''
	plane that is parallel to $B^r$ and
	containing the ``lower'' face of $B^t$.
%
%
        We adopt this latter $H_0$ to have a simpler geometric structure.

	This completes the description of $\wtFp(B)$.  It should
	be clear that checking if $\wtFp(B)$ intersects any feature $f$
	is relatively easy (since it is even a $\Pi_1$-set).
%
%
In Appendix~\ref{app-rod-pfs} 
we prove the following theorem:
\begin{theorem}
The approximate footprint $\wtFp(B)$ as defined for a rod robot
satisfies Eq.~\refeq{incl1}, i.e., there exists some fixed constant
$\sigma >1$ such that $\wtFp(B/\sigma) \ib \Fp(B) \ib \wtFp(B)$.
\end{theorem}
%
%
%
\remove{ 
\noindent
	Remark: a more accurate approximate footprint is given by
%
%
		$\Fp_{\,0}(B)\oplus Ball(B^t-m_B)$.
	Moreover, this is a $\Sigma_2$-set,
	written as the union of 5 balls of radius $r_B$
	with another somewhat complicated $\Pi_1$-set,
	and also 8 ``sleeves'' corresponding to 
	Minkowski sum of 8 edges of $\Fp_{\,0}(B)$ with $Ball(r_B)$.
	The 5 balls are the red balls $D_1\dd D_5$ shown in
	\refFig{rod-Minkowski-sum}.
	The four sleeves corresponding to straight edges are truncated
	cylinders; the other four sleeves corresponding to round
	edges are truncated rings.
	Clearly, this representation is much more complicated to implement
	than our $\wtFp(B)$.  This amounts to a (G2) versus (G3) trade-off.
} 

\sect[5]{Soft Predicates for a Ring Robot} \label{se-soft-ring}
\label{sect:ring}
	Let $R_0$ be a ring robot.  Its footprint
	is an embedded circle of radius $r_0$.
	First we show how to compute
		$\Sep(C,f)$,
	the separation of an embedded circle $C$ from a feature $f$.
	This was treated in detail by Eberly
	\cite{eberly:distance-to-circles}.
	This is easy when $f$ is a point or a plane.
	When $f$ is a line, Eberly gave two formulations:
	they reduce to solving a system of 2 quadratic
	equations in 2 variables, and hence to solving a quartic equation;
%
	see Appendix~\ref{appendixE-Separation}. 
	The predicate ``Does $f$ intersect $C\oplus Ball(r')$, a
	ring of thickness $r'$?'' is needed later; it 
	reduces to ``Is $\Sep(C,f) \le r'$?''.

	Our next task is to describe an approximate footprint,
	First recall the round cone of box $B$ defined in the previous section:
	$Cone(B) = Cone(m_B, Ball(m_B+B^r))$.
	Let $\theta=\theta(B)$ be the half-angle of this cone, and 
	$c$ the center of $B^r$.  Here, we think of $c$ as a point of
	$\sqs$, and define $\gamma(B)\as m_B\times c$
	viewed as an element of $\RR^3\times \sqs$.  Call
	$\gamma(B)$ the \dt{central configuration} of box $B$.
	Let $Ray(B)$ be the ray from $m_B$ through $m_B+c$.
	If $Plane(B)$ is the plane through $m_B$ and normal to $Ray(B)$,
	then the footprint $\Fp(\gamma(B))$ is an embedded
	circle lying in $Plane(B)$.
	We define the \dt{inner footprint} of $B$ as
		$\Fp_{\,0}(B) \as \Fp(m_B\times B^r).$
	The map $q\mapsto \ol{q}$ is the inverse of $q\mapsto \whq$,
	taking $c\in \sqs$ to $\ol{c}\in S^2$.
	It is hard to work with $\Fp_{\,0}(B)$.
	Instead consider the set $D(B)$ of all points in $S^2$
	whose distance\footnote{
	    Recall that $S^2$ is a metric space whose geodesics
	    are arcs of great circles.
	}
	from $\ol{c}$ is at most $\theta(B)$.
	So $D(B)$ is the intersection of $S^2$ with
	a round cone with ray from the origin to $c$.
	Then we have $\Fp_0(B)\ib\Fp_1(B)$ where
\beql{Fp_1(B)}
		\Fp_1(B) \as \Fp(m_B\times D(B)).
\eeql
%
          
	    	\begin{figure}[htb]
	    	  \begin{center}
		   \scalebox{0.28}{
	    	     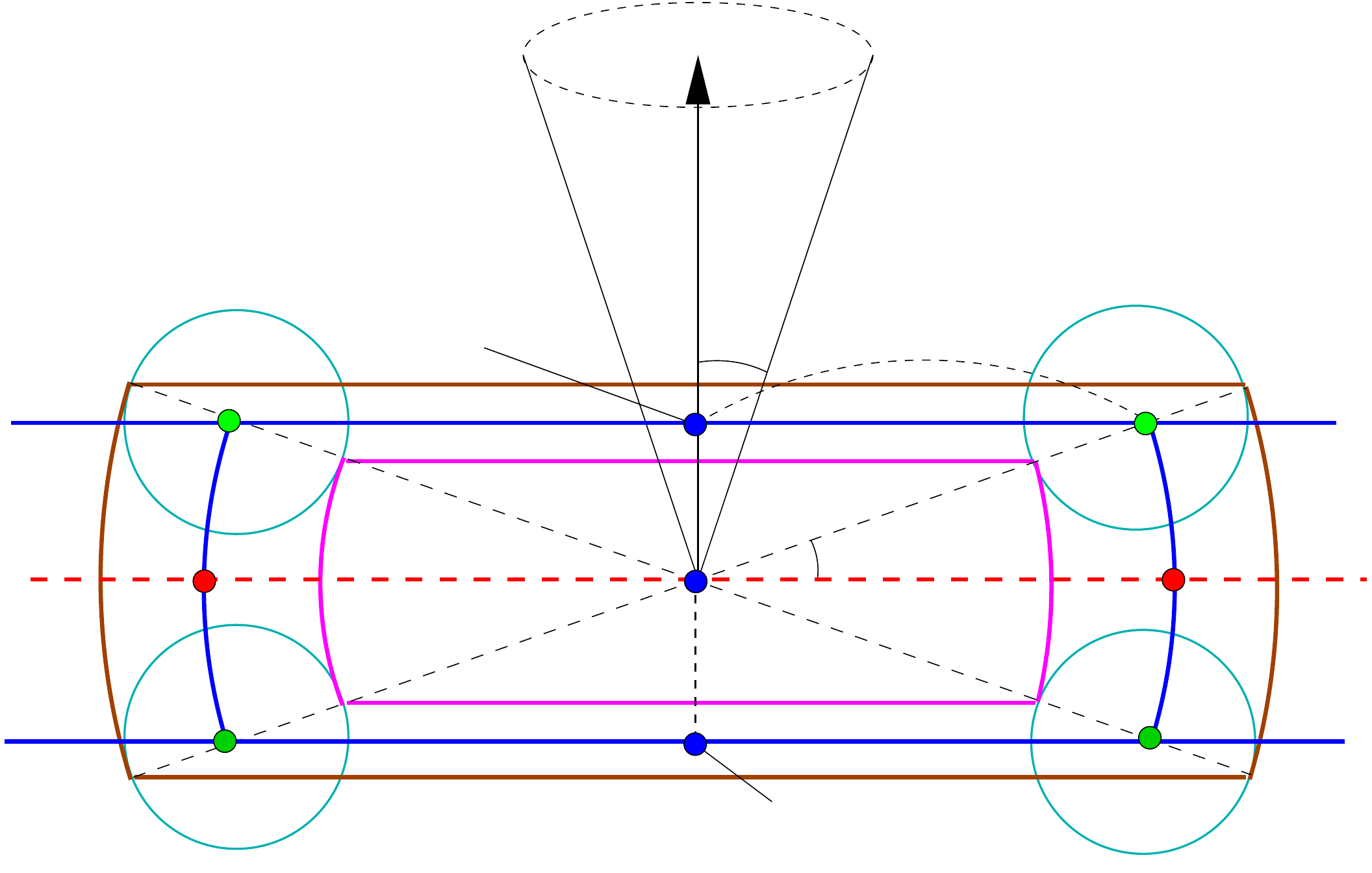}
	    	   \caption{Ring Robot:
		central cross-section of $\Fp_1(B)$ 
		appears as two blue arcs.  $\wtFp(B)$ equals the
		union of two ``thick rings'' and a ``truncated annulus''.
		The axis of $Cone(B)$ is shown as a vertical ray. 
                Each $Ball$ has radius $r_B$.            
		}
	    	   \label{fig:ring-Minkowski-new}
	    	  \end{center}
	    	\end{figure} 	
%
	Our main computational interest
	is the approximate footprint of $B$ defined as
\beql{wtFp(B)}
		\wtFp(B) \as \Fp_1(B)\oplus Ball(r_B).
\eeql
	Note that $\Fp_1(B)$ has a simple geometric description.
	We illustrate this in \refFig{ring-Minkowski-new} using
	a central cross-section with a plane through $m_B$
	containing the axis of $Cone(B)$ (the axis of $Cone(B)$
	is drawn vertically).  The footprint of $\gamma(B)$
	is a circle that appears as two red dots in the horizontal
	line (i.e., $Plane(B)$).
	Let $S^2(m_B,r_0)$ denote the 2-sphere centered at $m_B$
	with radius $r_0$.  Then $\Fp_1(B)$ is the intersection
	of $S^2(m_B,r_0)$ with a slab (i.e., intersection of
	two half-spaces whose
	bounding planes $P_1$ and $P_2$ are parallel to $Plane(B)$).
	These planes appear as two horizontal
	blue lines in \refFig{ring-Minkowski-new}.
	In the cross section, $\Fp_1(B)$ are seen as two blue circular arcs. 
	For $i=1,2$, let
	$C_i = P_i\cap S^2(m_B,r_0)$; it is an embedded circle that
	appears as a pair of green points in \refFig{ring-Minkowski-new}.
        Each $C_i$ is centered at
        $O_i$, with radius $r = r_0 \cos \theta$;
        see \refFig{ring-Minkowski-new}.

	We can now describe a $\Sigma_2$-decomposition of $\wtFp(B)$:
	it is the union of two ''thick rings'',
	$C_1\oplus Ball(r_B)$ and $C_2\oplus Ball(r_B)$
	(both of thickness $r_B$),
	and a shape $Ann(B)$ which
	we call a \dt{truncated annulus}.  First of all,
	the region bounded between the spheres $S^2(m_B,r_0+r_B)$
	(the brown arcs in the figure)
	and $S^2(m_B,r_0-r_B)$ (the magenta arcs)
	is called a (solid) annulus.
	Let $C^*_i$ denote the embedded disc whose relative boundary
	is $C_i$.  Then we have two round cones, $Cone(m_B,C^*_1))$ and
	$Cone(m_B,C^*_2))$.  Together, they form a
	{\em double cone} that is actually a simpler object for computation!
	Finally, define $Ann(B)$ to be the intersection of the
	annulus with the complements of the double cone.

For each thick ring $C_i\oplus Ball(r_B)$, deciding 
``Does a feature $f$ intersect $C_i\oplus Ball(r_B)$?'' 
is equivalent to ``Is $\Sep(C_i, f) \le r_B$?'' (see
beginning of this section). Appendix~\ref{appendixE-Separation} 
discusses this computation 	
and proves (in~\ref{appendixE-Proof}) 
the following theorem:
%
%
%
\begin{theorem}
The approximate footprint $\wtFp(B)$ as defined for a ring robot
satisfies Eq.~\refeq{incl1}, i.e., there exists some fixed constant
$\sigma >1$ such that $\wtFp(B/\sigma) \ib \Fp(B) \ib \wtFp(B)$.
\end{theorem}
%
%
\remove{ 
\begin{theorem} \label{thm-ring-2-properties}
$\wtFp(B)$ satisfies the inclusion properties in Eqs.~\refeq{incl1}
  and \refeq{incl2}. Namely:\\
(a) (Inheritance)
Let $B_1$ be a child of $B$. Then $\wtFp(B_1) \ib \wtFp(B)$.\\
(b) There exists some fixed constant $\sigma >1$ such that $\wtFp(B /
\sigma) \ib \Fp(B)$.
\end{theorem}
%
} 
	\ignore{
		We need to prove that $\wtFp(B/\sigma)\ib Fp_0(B)$.

		Show that 
			$\Fp_q(B/\sigma)\ib \Fp_0(B)$
		for some $\sigma$.
		Choose the $\sigma$ so that $Ball(r_B/\sigma)\ib B_t$.
		Then the result follows!
			
	}%

	\ignore{
	``A Discussion of the Cases when Two Quadratic Equations Involving Two
	Variables can be Solved by the Method of Quadratics''
		by Adelaide Denis,
		The American Mathematical Monthly
		Vol. 10, No. 8/9 (Aug. - Sep., 1903), pp. 192-199
	Another approach, from stack exchange,
	show how this can be reduced
	to a special case of $f(x,y)=g(x,y)=0$ where $f$ is general
	quadratic, but $g$ is bilinear (there are only $xy$ terms
	but not $x^2$ or $y^2$ terms).  This special form admits
	a simpler formula.
	}%

\sect[6]{Practical Efficiency of Correct Implementations} \label{se-expt}
%
        We have developed $\vareps$-exact planners 
        for rod and ring robots.
	We have explicitly exposed all the details necessary for
	a correct implementation, i.e., criterion (G1).
	The careful design of the
	approximate footprints of boxes as $\Sigma_2$-sets
	ensures (G2), i.e., it would be relatively easy to implement.
	We now address (G3) or practical efficiency.
	For robots with 5 or more DOFs, it becomes extremely critical
	that good search strategies are deployed.
        In this paper, we have found
	that some form of Voronoi heuristic is extremely effective:
the idea is to find paths along Voronoi curves
(in the sense of \cite{odun-yap:disc:85,odun-sharir-yap:vorI:86}),
and exploit subdivision Voronoi 
techniques based (again) on the method of features
\cite{yap-sharma-lien:vor:12,bennett-papadopoulou-yap:minimization:16}.
There are subtleties necessitating the use of pseudo-Voronoi curves
\cite{lee-choset:sensor-rod-planning:05,odun-sharir-yap:vorI:86,odun-sharir-yap:vorII:87}.
Since we do not rely on Voronoi heuristics for correctness,
simple expedients are available.
\remove{
Moreover, we could exploit the fact that we have also introduced
Voronoi diagram algorithms based on our method of features
\cite{yap-sharma-lien:vor:12,bennett-papadopoulou-yap:minimization:16}.
The basic idea is to try to move along the Voronoi curves which, in a
$d$-DOF robot, are defined generally by $d$ features.  This has many
difficulties including the inability to guarantee that such curves
exist (see
\cite{lee-choset:sensor-rod-planning:05,odun-sharir-yap:vorI:86,odun-sharir-yap:vorII:87}).
But we are not relying on Voronoi diagrams for our correctness, and
various expedients are available.
}
	To recognize Voronoi curves, we
	maintain (in addition to the collision-detection
	feature set $\wtphi(B)$), the \dt{Voronoi feature set}
	$\wtphi_V(B)$.  These two sets have some connection
	but there are no obvious inclusion relationships. 

%
\FigEPS{ring-trace-subdiv-rand40}
		   { 0.25 } 
		  {Ring robot amidst 40 random tetrahedra:
                   (a) trace of a found path; 
                   (b) subdivision of translational boxes on the path.
                  } 
%

 \FigEPS{ring-trace-sub-posts}
                   {0.33}
		  {Ring robot amidst 
%
%
                   pillars and L-shaped posts (Posts):
                   (a) trace of a found path; 
                   (b) subdivision of translational boxes on the path.
                  } 

 \FigEPS{ring-no-path-sub-posts2}
                   {0.33}
		  {Ring robot amidst 
%
%
                   another set of pillars and posts (Posts2):
                   (a) start and goal configurations (no path found); 
                   (b) subdivision of translational boxes during the search.
                  } 

%
%
%
%
    \begin{center}
	\begin{table*}[htbp]
\hspace{-1.2em}
            \makebox[\textwidth][c]{\includegraphics[width=1.08\textwidth]{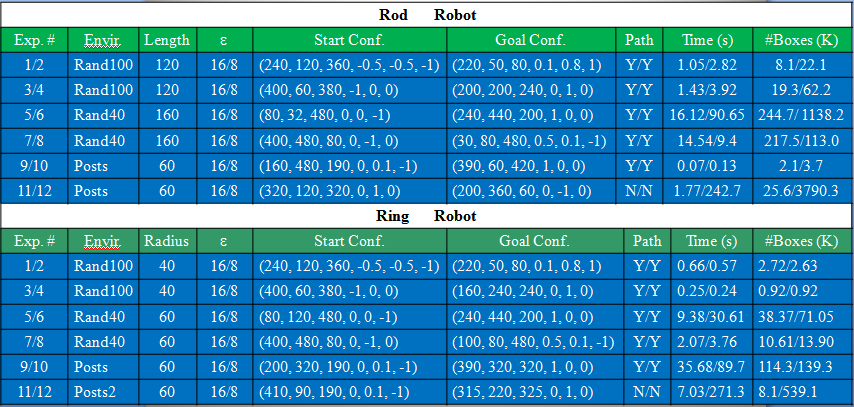}} 
 
	\caption{Rod and Ring Experiments. 
%
%
	}
	\label{fig:table1}
	\end{table*}
	\end{center}
%
%
%
	
	Our current implementation achieves
        near real-time performance
	(see video  \\
	\myHrefx{http://cs.nyu.edu/exact/gallery/rod-ring/rod\_ring.html}).
	Table~\ref{fig:table1} summarizes experiments on our rod and ring
	robots. 
	The environments Rand100, Rand40 (100 and 40 random tetrahedra), 
        Posts and Posts2 are shown in 
%
%
        Figs.~1,~5,~6 and~7. 
%
%
%
%
%
The dimensions of the environments are $512^3$. 
%
%
	Our implementation uses {\tt C++} and OpenGL on the
	Qt platform.  Our code, data and experiments are
	distributed\footnote{ 
		{http://cs.nyu.edu/exact/core/download/core/.}
	        }
	with our open source \corelib.
        We ran our experiments on a MacBook Pro 
        under Mac OS X 10.10.5 
        with a 2.5 GHz Intel Core i7 processor, 
        16GB DDR3-1600 MHz RAM and 500GB Flash Storage.
	Details about these experiments are found
	in a folder in \corelib\ for this paper;
	a {\tt Makefile} there can automatically run all the experiments.
	Thus these results are reproducible from the data there.  

	Table~\ref{fig:table2} (correlated with Table~\ref{fig:table1} by the Exp \#'s)
        compares our methods with various
        sampling-based planners in OMPL~\cite{ompl}, where we accepted
        the default parameters and each instance was run 10 times,       
        with the ``average time (in s)/standard deviation/success rate''
        reported.  This comparison has various caveats: we simulated
        the rod and ring robots by polyhedral approximations.
%
%
        We usually outperform RRT in cases of PATH.  In case of
        NO-PATH, we terminated in real time while all sampling methods
        timed out (300s).

%
%
    \begin{center}
	\begin{table*}[htbp]
\hspace{-0.5em}
            \makebox[\textwidth][c]{\includegraphics[width=1.14\textwidth]{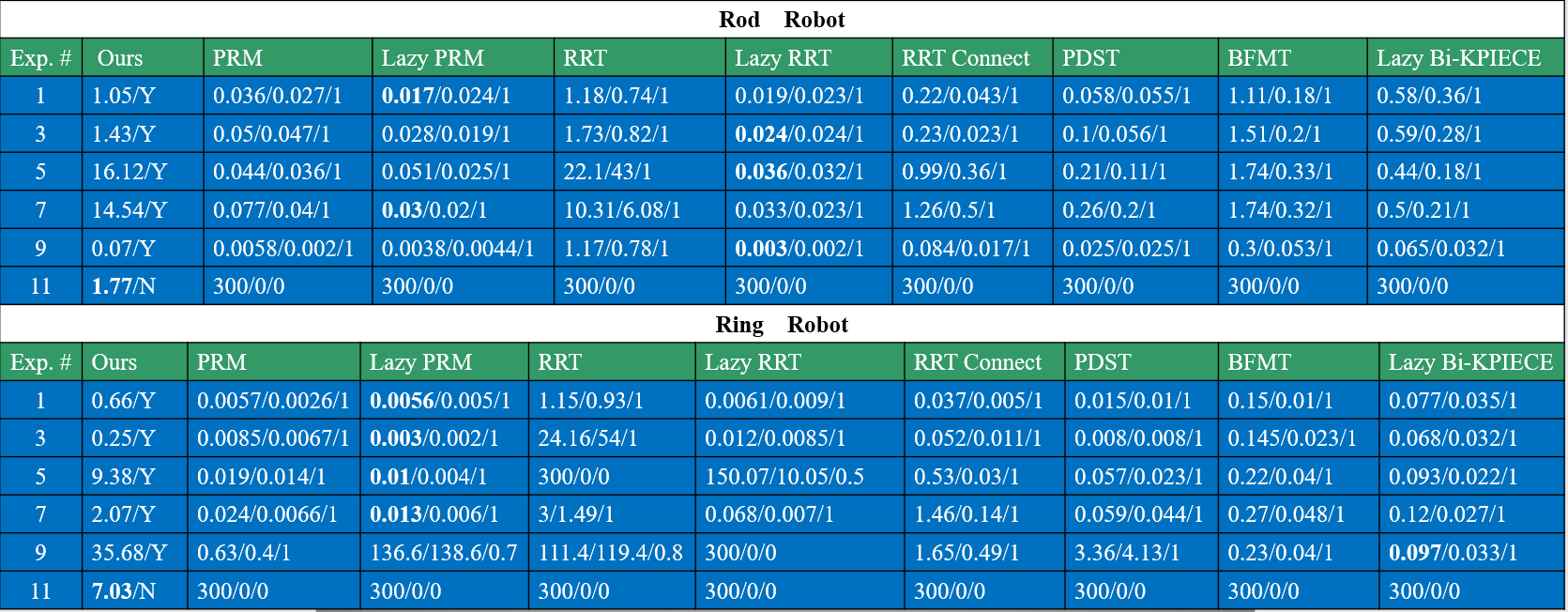}} 
	\caption{Comparison with Sampling Methods in OMPL 
                 (the best run-time is shown in bold).
%
%
	}
	\label{fig:table2}
	\end{table*}
	\end{center}
%
%
\remove{
	\noindent {\bf Correct Implementation of Soft Exact Algorithms}\\
	We have provided
	%
	%
	an ``exact'' description of planners for a rod and a ring, albeit a
	``soft kind''.
	%
	%
	We claim that all our computations can be guaranteed in the soft
	sense.
	This is possible because all the inequalities in our algorithms are
	``one-sided'' in the sense that we do not assume that the failure of
	an inequality test implies the complementary condition (as in exact
	(unqualified) computation).
	More detailed discussions are in Appendix~\ref{app-implement}.  
}%

%
\remove{ 
%
\ssect{Correct Implementation of Soft Exact Algorithms}
	We have provided 
%
%
        an ``exact'' description of planners
	for a rod and a ring, albeit a ``soft kind''
	that admits a user-controlled amount of numerical indeterminacy.
	The reader may have noticed
	that we formulated precise mathematical relations
	and exact geometric shapes for which various inclusions
	must be verified for correctness.   Purely numerical
	computations (even with arbitrary precision)
	cannot ``exactly determine'' such relations in general.
	Nevertheless,
	we claim that all our computations can be guaranteed
	in the soft sense.  The basic idea is that for
	each box $B$, all the computations
	associated with $B$ is computed to some absolute
	error bound that at most $r_B/K^*$ where $r_B$
	is the box radius and $K^*$ is a constant depending on the
	algorithm only.  Thus, as
	boxes become smaller, we need higher precision
	(but the resolution $\vareps$ ensures termination).
	Moreover, the needed precision requires no special programming
	effort.
	
	This is possible because all the inequalities in our
	algorithms are ``one-sided'' in the sense that
	we do not assume that the failure of an inequality test implies
	the complementary condition (as in exact (unqualified) computation).
	We can define a
	\dt{weak feature set} denoted $\whphi(B)$ with this property:
		\[\whphi(B/\sigma)\ib \wtphi(B)\ib \whphi(B)\]
	for some $\sigma>1$.
	The ``weak'' $\whphi(B)$ is
	not uniquely determined (i.e., $\whphi(B)$ can be {\em any} set
	that satisfies the inequalities).  
	In contrast, the set $\wtphi(B)$ is mathematically precise and
	unique.  If we use $\whphi(B)$ instead of $\wtphi(B)$,
	the correctness of our planner remains intact. 
	Moreover, the weak set
	$\whphi(B)$  can be achieved as using numerical approximation
	(note: we do not need "correct rounding" from our bigFloats,
	so \gmp\ suffice).
} 

\sect[7]{Conclusions}
%
	Path planning in 3D has many challenges.
	Our 5-DOF spatial robots have pushed the current limits of
	subdivision methods.  To our knowledge there is no similar
	algorithm with comparable rigor or guarantees.
	Conventional wisdom says
	that sampling methods can achieve higher DOFs than
	subdivision.  By an estimate of
	Choset et al \cite[p.~202]{choset-etal:bk},
	sampling methods are limited to $5-12$ DOFs.
	We believe our approach can reach 6-DOF spatial robots.
	Since resolution-exactness delivers stronger
	guarantees than probabilistic-completeness,
	we expect a performance hit compared to sampling methods.  
	But for simple planar robots (up to 4
	DOFs)~\cite{wang-chiang-yap:motion-planning:15,luo-chiang-lien-yap:link:14,yap-luo-hsu:thicklink:16,zhou-chiang-yap:complex-robot:18}
%
%
	we observed no such trade-offs because we
	outperform state-of-the-art sampling methods
	(such as OMPL \cite{ompl})
	often by two orders of magnitude.
	But in the 5-DOF robots of this paper, we see that
	our performance is competitive with sampling methods.
%
%
%
	It is not clear to us that subdivision
	is inherently inferior to sampling (we can also do random subdivision).
	It is true that
	each additional degree of freedom is conquered only with
	effort and suitable techniques.
	This remark seems to cut across both subdivision
	and sampling approaches; but it hits subdivision harder
	because of our stronger guarantees.

\bibliographystyle{abbrv} 

\bibliography{test,st,yap,exact,geo,alge,math,com,rob,cad,algo,visual,gis,quantum,mesh,tnt,fluid} 

\newpage
\appendix
\section*{Appendices} 
%
%
  In the following appendices,
  the figure numbers are continued from the paper.
%
\section{Appendix: Elements of Soft Subdivision Search} \label{app-SSS}
%
%
	  We review the the notion of soft predicates and how
	  it is used in the SSS Framework.
	  See \cite{wang-chiang-yap:motion-planning:15,sss,luo-chiang-lien-yap:link:14}
	  for more details.
	  
	  \subsection{Soft Predicates}
	  The concept of a ``soft predicate'' is relative to some exact
	  predicate.  
	  Define the exact predicate $C:\cspace\to \set{0,+1,-1}$ where 
	  $C(x)=0/+1/-1$ (resp.) if configuration $x$ is semi-free/free/stuck.
	  The semi-free configurations are those on the boundary of $\cfree$.
	  Call $+1$ and $-1$ the \dt{definite values}, and $0$ the
	  \dt{indefinite value}.
	  Extend the definition to any set $B\ib\cspace$:
	  for a definite value $v$, define $C(B)=v$ iff $C(x)=v$ for all $x$. 
	  Otherwise, $C(B)=0$.  
	  Let $\intbox(\cspace)$ denote the set of $d$-dimensional boxes in
	  $\cspace$.  A predicate $\wtC:\intbox(\cspace)\to\set{0,+1,-1}$ is a
	  \dt{soft version of $C$} if
	  it is conservative and convergent.  \dt{Conservative} means that
	  if $\wtC(B)$ is a definite value, then $\wtC(B)=C(B)$.
	  \dt{Convergent} means that if for any sequence $(B_1,B_2,\ldots)$ of
	  boxes, if $B_i\to p\in\cspace$ 
	  as $i\to\infty$, then $\wtC(B_i)=C(p)$ for $i$ large enough. 
	  To achieve resolution-exact algorithms, we must ensure
	  $\wtC$ converges quickly in this sense:
	  say $\wtC$ is \dt{effective} if there is a constant
	  $\sigma>1$ such if $C(B)$ is definite, then $\wtC(B/\sigma)$ is
	  definite.
	  
	  \subsection{The Soft Subdivision Search Framework}
	  An SSS algorithm maintains
	  a subdivision tree $\TTT=\TTT(B_0)$ rooted at a given box $B_0$.
	  Each tree node is a subbox of $B_0$.
	  We assume a procedure $\splittt(B)$ that subdivides a given leaf box
	  $B$ into a bounded number of subboxes which becomes the children of
	  $B$ in $\TTT$.  Thus $B$ is ``expanded'' and no longer a leaf.
	  For example, $\splittt(B)$ might create
	  $2^d$ congruent subboxes as children.  Initially $\TTT$ has just
	  the root $B_0$; we grow $\TTT$ by repeatedly expanding its leaves.
	  The set of leaves of $\TTT$ at any moment constitute a subdivision of
	  $B_0$.
	  Each node $B\in \TTT$ is classified using a soft predicate $\wtC$ 
	  as $\wtC(B)\in\set{\mixed,\free,\stuck}=\set{0,+1,-1}$. 
	  Only \mixed\ leaves with radius $\ge\vareps$ are candidates for
	  expansion.
	  We need to maintain three auxiliary data structures:
	  \bitem
	  \item
	  A priority queue $Q$ which contains all candidate boxes.
	  Let $Q.\getnext()$ remove the box of highest priority from $Q$.
	  The tree $\TTT$ grows by splitting $Q.\getnext()$.
	  \item
	  A \dt{connectivity graph} $G$ whose nodes are the \free\ leaves in
	  $\TTT$,
	  and whose edges connect pairs of boxes that are adjacent,
	  i.e., that share a $(d-1)$-face.
	  \item
	  A Union-Find data structure for connected components of $G$.
	  After each $\splittt(B)$, we update $G$ and insert new
	  \free\ boxes into the Union-Find data structure and perform
	  unions of new pairs of adjacent \free\ boxes.
	  \eitem
	  
	  Let $Box_{\TTT}(\alpha)$
	  denote the leaf box containing $\alpha$ (similarly for
	  $Box_{\TTT}(\alpha)$).  
	  The SSS Algorithm has three WHILE-loops.  The first WHILE-loop will
	  keep splitting $Box_{\TTT}(\alpha)$ until it becomes \free, or declare
	  NO-PATH when $Box_{\TTT}(\alpha)$ has radius less than $\vareps$.
	  The second WHILE-loop does the same for $Box_{\TTT}(\beta)$.
	  The third WHILE-loop is the main one: it will keep splitting
	  $Q.\getnext()$ until a path is detected or $Q$ is empty.
	  If $Q$ is empty, it returns NO-PATH.  Paths are detected when
	  the Union-Find data structure tells us that
	  $Box_{\TTT}(\alpha)$ and $Box_{\TTT}(\beta)$ are in the same
	  connected component.
	  It is then easy to construct a path.  Thus we get:
	    \progb{
              \lline[-1]{\sc SSS Framework:}
		\lline[0] \INPUt:  Configurations $\alpha,\beta$, tolerance
			$\vareps>0$, box $B_0\in \cspace$.
		\lline[0] \OUTPUt: Path from $\alpha$ to $\beta$ in
			$\FP(R_0,\Omega)\cap B_0$.
		  \lline[5] Initialize a subdivision tree $\TTT$ with root
		  	$B_0$.
	          \lline[5] Initialize $Q, G$ and union-find data structure.
	          \lline[5] While ($Box_{\TTT}(\alpha) \neq \free$)
		    \lline[10] If radius of $Box_{\TTT}(\alpha))$ is $<
		    	\vareps$, Return(NO-PATH)
	            \lline[10] Else \splittt($Box_{\TTT}(\alpha))$
	          \lline[5] While ($Box_{\TTT}(\beta) \neq \free$)
		    \lline[10] If radius of $Box_{\TTT}(\beta))$ is $<
		    	\vareps$, Return(NO-PATH)
	            \lline[10] Else \splittt($Box_{\TTT}(\beta))$
	          \lline[5] \commenT{MAIN LOOP:}
		  \lline[5] While ($Find(Box_{\TTT}(\alpha))\neq
		  	Find(Box_{\TTT}(\beta))$)
	            \lline[10] If $Q_{\TTT}$ is empty, Return(NO-PATH)
	            \lline[10] $B\ass Q_{\TTT}.\getnext()$
	            \lline[10] $\splittt(B)$
		  \lline[5] Generate and return a path from $\alpha$ to $\beta$
		  	using $G$.
	    }

\ignore{  
\progb{
	    \lline[-5] {\sc Split($B$)}
	      \lline[0] \INPUt: Box $B$
	        \lline[5] If ($\phi_0(B)$ is empty)
	          \lline[10] Do T-split.
	        \lline[5] Else
	          \lline[10] If ($B^r$ is equal to $\sqs$)
	            \lline[15] Do 6-way R-splits.
	          \lline[10] Else if($r_B > \vareps$)
	            \lline[15] Do T-split.
	          \lline[10] Else
	            \lline[15] Do 4-way R-splits.
	        \lline[5] For each $B'\in B.children$
	          \lline[10] Classify $B'$.
}
	
\progb{
	  \lline[-5] {\sc Soft Classification of Box $B$}
	  \lline[0] \INPUt: Box $B$
	  \lline[0] \OUTPUt: Status $s \in {\stuck/\mixed/\free}$
	      \lline[5] Initialize $\wtphi_i(B)$ is empty ($i=0,1,2$).
	      \lline[5] For each $C\in \wtphi_0(B.parent)$
	      \lline[10] If ($\Sep_C(m_B)\le\ell_0+r_B$ and $\wh{C}\ib B^r$)
	          \lline[15] Add $C$ to $\wtphi_0(B)$.
	      \lline[5] For each $E\in \wtphi_1(B.parent)$
		\lline[10] Compute the projection $\wh{E}$ of $E$ onto
		$\wh{S^2}$.
		\lline[10] If($\Sep_E(m_B)\le\ell_0+r_B$ and $\wh{E} \cap B^r
		\neq \emptyset$)
	          \lline[15] Add $E$ to $\wtphi_1(B)$.
	      \lline[5] For each $W\in \wtphi_2(B.parent)$
		\lline[10] Compute the closest point $c$ between wall $W$ and
		$m_B$.
		\lline[10] Compute the projection of $B^r$ onto plane $\ol{W}$
		and obtain a quadrilateral $Q$.
		\lline[10] The circle $D$ centered at $c$ is formed by the rod
		on plane $\ol{W}$.
		\lline[10] If($\Sep_W(m_B)\le\ell_0+r_B$ and ($\wh{c}\ib B^r$
		or $Q$ intersects $D$))
	          \lline[15] Add $W$ to $\wtphi_2(B)$.
	      \lline[5] If($\wtphi(B)$ is empty)
	        \lline[10] If $B.parent$ is NULL, Return $\mixed$
	        \lline[10] Find feature $f\in \wtphi(B.parent)$ that is closest to $m(B)$
	        \lline[10] If($f$ is a wall with equation $L$)
	          \lline[15] Return the sign of $L(m(B))$ where $-1/0/+1$ corresponds to $\stuck/\mixed/\free$ (respectively).
	        \lline[10] Else If($f$ is an edge)
	          \lline[15] If(is a concave edge) \Return $\stuck$
	          \lline[15] Else, Return $\free$ ($f$ is a convex edge)
	        \lline[10] Else ($f$ is a corner)
	          \lline[15] If($f$ is a convex corner), Return $\free$
	          \lline[15] Else If($f$ is a concave corner), Return $\stuck$
	          \lline[15] Else $f$ is a mixed corner
	            \lline[20] Find a wall $w'$ with equation $L'$ incident to $f$
	            \lline[20] and has the minimum angle with plane $\Pi: \overrightarrow{m(B)-f} \cdot (x - f) = 0 $
	            \lline[20] Return the sign of $L'(m(B))$
	      \lline[5] Else, Return $\mixed$
}
	
\begin{algorithm}[H]
	  \caption{Split Box $B$}
	  \label{alg::Split}
	  \begin{algorithmic}[1]
	    \Require{Box $B$}
	    \Ensure{$B.children$}
	    \If{$\phi_0(B)$ is empty}
	      \State Do T-split. 
	    \Else
	      \If{$B^r$ is equal to $\sqs$}
	        \State Do 6-way R-splits.
	      \ElsIf{$r_B > \vareps$}
	        \State Do T-split.
	      \Else
	        \State Do 4-way R-splits.
	      \EndIf
	    \EndIf
	    \For{each $B'\in B.children$}
	      \State Classify $B'$.
	    \EndFor
	  \end{algorithmic}
	  \end{algorithm}
	
	  \begin{algorithm}[H]
	  \caption{Soft Classification of Box $B$}
	  \label{alg::SoftClassification}
	  \begin{algorithmic}[1]
	    \Require{Box $B$}
	    \Ensure{Status $s \in{\stuck/\mixed/\free}$)}
	      \State Initialize $\wtphi_i(B)$ is empty ($i=0,1,2$).
	      \For{each $C\in \wtphi_0(B.parent)$}
	        \If{$\Sep_C(m_B)\le\ell_0+r_B$ and $\wh{C}\ib B^r$}
	          \State Add $C$ to $\wtphi_0(B)$.
	        \EndIf
	      \EndFor
	      \For{each $E\in \wtphi_1(B.parent)$}
	        \State Compute the projection $\wh{E}$ of $E$ onto $\wh{S^2}$.
	        \If{$\Sep_E(m_B)\le\ell_0+r_B$ and $\wh{E} \cap B^r \neq \emptyset$}
	          \State Add $E$ to $\wtphi_1(B)$.
	        \EndIf
	      \EndFor
	      \For{each $W\in \wtphi_2(B.parent)$}
	        \State Compute the closest point $c$ between wall $W$ and $m_B$.
	        \State Compute the projection of $B^r$ onto plane $\ol{W}$ and obtain a quadrilateral $Q$.
	        \State The circle $D$ centered at $c$ is formed by the rod on plane $\ol{W}$.
	        \If{$\Sep_W(m_B)\le\ell_0+r_B$ and ($\wh{c}\ib B^r$ or $Q$ intersects $D$)}
	          \State Add $W$ to $\wtphi_2(B)$.
	        \EndIf
	      \EndFor
	      \If{$\wtphi(B)$ is empty}
	        \If{$B.parent$ is NULL} \Return $\mixed$
	        \EndIf
	        \State Find feature $f\in \wtphi(B.parent)$ that is closest to $m(B)$
	        \If{$f$ is a wall with equation $L$}
	          \State \Return the sign of $L(m(B))$ where $-1/0/+1$ corresponds to $\stuck/\mixed/\free$ (respectively).
	        \ElsIf{$f$ is an edge}
	          \If{$f$ is a concave edge} \Return $\stuck$
	          \Else{ \Return $\free$ ($f$ is a convex edge)}
	          \EndIf
	        \Else \Comment{$f$ is a corner}
	          \If{$f$ is a convex corner} \Return $\free$
	          \ElsIf{$f$ is a concave corner} \Return $\stuck$
	          \Else \Comment{$f$ is a mixed corner}
	            \State Find a wall $w'$ with equation $L'$ incident to $f$
	            \State and has the minimum angle with plane $\Pi: \overrightarrow{m(B)-f} \cdot (x - f) = 0 $
	            \State \Return the sign of $L'(m(B))$
	          \EndIf
	        \EndIf
	      \Else
	        \State \Return $\mixed$
	      \EndIf
	  \end{algorithmic}
\end{algorithm}
}%
	  
	  See \cite{sss} for the correctness of this
	  framework under very general conditions.
	  Note that $Q$ is a priority queue, and $Q.\getnext()$
	  extracts a box of lowest priority.
	  The correctness of our algorithm does not depend on choice
	  of priority.  E.g., we could have
	  randomly-generated priority
	  to simulate some form of random sampling.
	  However, choosing a good priority can have a great impact on
	  performance.
	  In our implementations, especially in 3-D,
	  we have found that heuristics based on Greedy Best-First
	  and some Voronoi heuristics are essential
	  for real-time performance.

	
	
	
	
	

\section{Appendix: Properties of Square Models, Classifying a Box, 
                   and Properties of $\wtphi'(B)$ }
\subsection{Proof: Properties of Square Models} \label{appendixC-square-model}
{\bf Lemma~1.}
{\em
$C_0=\sqrt{3}$.
}\\
%
	\bpf Let $B$ be the ball whose boundary is $S^2$ and $C=[-1,1]^3$.
	Then $B\ib C \ib \sqrt{3}B$.   From any geodesic $\alpha$ of $S^2$,
	we obtain a corresponding geodesic $\alpha'$ on the surface of
	$\sqrt{3}B$, and a geodesic $\wh{\alpha}$ of $\sqs=\partial(C)$. 
	Observe
	that $|\alpha| \le |\wh{\alpha}| \le |\alpha'|$ where $|\cdot|$
	is the length of a geodesic.  But $|\alpha'| =\sqrt{3}|\alpha|$.
	This proves that
	$1\le \frac{|\wh{\alpha}|}{|\alpha|} \le \sqrt{3}$,
	i.e., $C_0\le \sqrt{3}$.
	This bound on $C_0$ is tight because for geodesic
	arcs in arbitrarily small neighborhoods of the corners of 
	$\sqs$, the bound is arbitrarily close to $\sqrt{3}$.
	\epf

\subsection{Classifying a Box}\label{appendixC-classify-box}
In Sec.~\ref{se-let-us-design} we mentioned using soft predicates
based on the ``method of features''
\cite{wang-chiang-yap:motion-planning:15} to classify a box
$B$. Recall that we classify $B$ as \mixed\ when the feature set is
non-empty; otherwise, we classify $B$ as \free\ or \stuck.
%
%
%
Now we discuss how to classify 
%
%
$B$ as \free\ or \stuck\ when its feature set is empty.
Suppose $\Omega$ is given as the union of a set of polyhedra
%
%
that may overlap (this situation arises in Sec.~\ref{se-expt}). 
Let $B'$ be the parent of $B$, then the feature set $\wtphi(B')$ is
non-empty. For each obstacle polyhedron $P$ in $\wtphi(B')$, we find
the feature $f \subseteq \partial P$ closest to $m_B$ and use $f$ to
decide whether $m_B$ is outside $P$. Then $m_B$ is outside $\Omega$
(and $B$ is \free) iff $m_B$ is outside all such polyhedra $P$. 

\remove{ 
\noindent {\bf ***??? YJC: Note: $m_B$ must be outside *all* such
  polyhedra $P$. So for *each* $P$ we must find the closest feature
  $f$ of $P$ to $m_B$, rather than just the *one* closest feature $f$
  of $\Omega$.} \\
} 

To find the feature $f \subseteq \partial P$ closest to $m_B$, we
first find among the corners of $P$ the one $f_c$ that is the closest.
Then among the edges of $P$ incident on $f_c$, we check if there exist
edges $e$ that are even closer (i.e., $\Sep(e, m_B) < \|f_c - m_B\|$ with
$\Sep(e, m_B)=\|p-m_B\|$ for some point $p$ interior to $e$) and if so pick
the closest one $f_e$. Finally, if $f_e$ exists, we repeat the process
for faces of $P$ incident on $f_e$ and pick the closest one $f_w$ (if
it exists). The closest feature $f$ is set to $f_c$ then updated to
$f_e$ and to $f_w$ accordingly if $f_e$ (resp.\ $f_w$) exists.

Given the feature $f \subseteq \partial P$ closest to $m_B$, we can
easily determine if $m_B$ is interior or exterior of $P$ when $f$ is a
wall or an edge.  When $f$ is a corner, it is slightly more
involved.  
%
%
We will classify a corner $f$ to be \dt{pseudo-convex} (resp.,
\dt{pseudo-concave}) if there exists a closed half space $H$ such that
(1) $f \in \partial H$, 
and (2) for any small enough ball $\Delta$ centered at $f$, we
have that $(H\cap P\cap \Delta) =f$ (resp.,
$H\cap\Delta \ib P\cap \Delta$).  
%
%
Note that if $f$ is locally convex (resp., locally
concave) then it is pseudo-convex (resp., pseudo-concave).
We call a corner $f$ an \dt{essential corner} if for all balls
$\Delta$ centered at $f$, $\Delta \cap \partial P$ is not a planar
set.  We may assume that our corners are essential; as consequence, no
corner can be both pseudo-convex and pseudo-concave.  However, it
is possible that a corner is neither pseudo-convex nor pseudo-concave;
we call such corners \dt{mixed}.
%
%
The lemma below enables us to avoid the difficulty of mixed corners.
\blem
Let $q\notin\partial P$ and $C$ a corner of $P$.
If $C$ is the point in $\partial P$ closest to $q$,
i.e., $\Sep_{\partial P}(q)=\|q-C\|$, then $C$ is either
pseudo-convex or pseudo-concave.  Hence $C$ cannot be a mixed corner.
Moreover, $q \in P$ iff $C$ is pseudo-concave.
\elem
\bpf
%
Let $\Delta$ be the ball centered at $q$ with radius $\|q-C\|$.
Since $\Sep_{\partial P}(q)=\|q-C\|$, we have $\Delta \cap \partial P =\set{C}$.
Let $H$ be the closed half-space such that $\partial H$
is tangential to $\Delta$ at the point $C$, and $q\notin H$.
This $H$ is a witness to either the 
pseudo-convexity or pseudo-concavity of $C$.  
In particular, $C$ is pseudo-concave iff $q\in P$.
\epf

%

\subsection{Proof: Properties of $\wtphi'(B)$} \label{app-pf-lemma2}
{\bf Lemma~2.}
{\em
If the approximate footprint $\wtFp(B)$ satisfies Eq.~\refeq{incl1},
%
%
%
then $\wtphi'(B)$ satisfies Eq.~\refeq{incl2}, i.e.,
	\[\wtphi'(B/\sigma) \ib \phi(B) \ib \wtphi'(B).\]
} 
%
\bpf  
Let $B$ be an aligned box. 
Define $\wtFp'(\cdot)$ recursively as follows:
(I) for an aligned box $B$, 
$\wtFp'(B) \as \wtFp(B)$ if
$B$ is the root, and $\wtFp'(B)\as$ 
$ \wtFp'(parent(B)) \cap \wtFp(B)$
otherwise;
(II) for a non-aligned box $B/\sigma$,
$\wtFp'(B/\sigma) \as \wtFp(B/\sigma)$ if
$B$ is the root, and $\wtFp'(B/\sigma)\as$ \\
$ \wtFp'(parent(B)/\sigma) \cap \wtFp(B/\sigma)$
otherwise.
Comparing with the recursive definitions of $\wtphi'(B)$ and of $\wtphi'(B/\sigma)$
(Eqs.~\refeq{wtphi'} and~\refeq{wtphi''}), it is easy to verify that
$\wtphi'(B) = \set{f\in\Phi(\Omega): f\cap \wtFp'(B)\neq\es}$,
and that $\wtphi'(B/\sigma) = 
\set{f\in\Phi(\Omega): f\cap \wtFp'(B/\sigma)\neq\es}$.
Therefore, we will show that $\wtFp'(B)$
satisfies Eq.~\refeq{incl1}, i.e., $\wtFp'(B/\sigma) \ib \Fp(B) \ib
\wtFp'(B)$, which implies that $\wtphi'(B)$ satisfies
Eq.~\refeq{incl2}. \\
%
%
%
%
(i) The case when $B$ is the root is easy.  Since
$\wtFp'(B) = \wtFp(B)$ and $\wtFp'(B/\sigma) = \wtFp(B/\sigma)$,
and also $\wtFp(B)$ satisfies Eq.~\refeq{incl1}, i.e.,
$\wtFp(B/\sigma) \ib \Fp(B) \ib \wtFp(B)$, we have
$\wtFp'(B/\sigma) = \wtFp(B/\sigma) \ib \Fp(B) \ib \wtFp(B) = \wtFp'(B)$,
as desired. \\
(ii) Now suppose $B$ is not the root. We proceed the proof in two
parts below.\\
(A) First we prove that $\Fp(B) \ib \wtFp'(B)$.
By definition, we have
$\wtFp'(B) = \wtFp'(parent(B)) \cap \wtFp(B)$.
Since $\wtFp(B)$ satisfies Eq.~\refeq{incl1}, $\wtFp(B)$ is a superset
of $\Fp(B)$.
Therefore it suffices to show that $\wtFp'(parent(B))$ is a superset
of $\Fp(B)$. But $\wtFp'(parent(B))$ is a superset of $\Fp(parent(B))$
(initially
for $parent(B)$
at the root and inductively going down), which in term is a superset
of $\Fp(B)$. \\
(B) Finally we prove that $\wtFp'(B/\sigma) \ib \Fp(B)$.
Since $\wtFp(B)$ satisfies Eq.~\refeq{incl1}, we have
$\wtFp(B/\sigma) \ib \Fp(B)$.
But $\wtFp'(B/\sigma) = \wtFp(B/\sigma) \cap \wtFp'(parent(B)/\sigma)$
is a subset of 
$\wtFp(B/\sigma)$ and hence the statement is true.
\epf
%

\remove{
Note: Using the approach (def. of pseudo-convexity/pseudo-concavity & the Lemma & Pf above),
      we do NOT need to worry about mixed corners, i.e.,
      In Algorithm 4 (in the appendix of scg2017 write-up (abs.tex there)), 
      when the closest feature is a corner, it will NEVER be a mixed corner.

This Algorithm 4 is NO Longer Needed to be presented in this paper!

For fixing the code, Algorithm 4 needs the following changes for this part.

Changes Needed:

1. Change the conditions of convex/concave corner to
   pseudo-convex/pseudo-concave corner.

2. Remove the case of mixed corner.

3. $f$ being pseudo-convex/pseudo-concave does not give 
   \free/\stuck\ to the box $B$ yet; it only decides that 
   $m_B$ is outside/inside the current polyhedron $P$. 

** Here is the portion of Algorithm 4 (copied here), the case of closest feature being a corner:
%
\noindent         
        {\tt //} This is the case where $f$ is a corner \\
%
           {\bf if} $f$ is a pseudo-convex corner {\bf return} $\free$ \\  
%
           {\bf elseif} $f$ is a pseudo-concave corner {\bf return} $\stuck$ \\ 
%
\remove{
           {\bf else} {\tt //} $f$ is a mixed corner \\
%
%
            Find a wall $w'$ with equation $L'$ incident to $f$
            and has the minimum angle with plane $\Pi: \overrightarrow{m(B)-f} \cdot (x - f) = 0 $\\
%
            {\bf return} the sign of $L'(m(B))$ \\
}
%
%
} 
%
%


\section{Appendix: Soft Predicate for a Rod --- Proofs} \label{app-rod-pfs}
%
%
\ignore{
In this section, we consider the robot $R_0$ to be a rod, which is a
line segment with one endpoint $A$ and length $r_0$. We typically
specify its configuration by the position of $A$ and the orientation
of the line segment. Let $B = B^t \times B^r$ be a subdivision box,
where $B^t$ is the translational box centered at $m_B$ with radius
$r_B$ and $B^r$ is the rotational box. A major issue in the design of
our soft predicate is to consider $\Fp(B)$, the (exact) footprint of
$R_0$ when moving around inside the box $B$, and construct its
desirable superset $\wtFp(B)$.  Recall that our soft predicate
$\wtC(B)$ depends on $\wtphi(B)$, the soft version of the feature
set. In order to construct $\wtphi(B)$ efficiently, for each type of
features $f$ (corner, edge, wall), we need to be able to decide
efficiently whether $f$ intersects $\wtFp(B)$.

First, consider $\Fp(m_B, B^r)$, the (exact) footprint of the rod
with $A$ at $m_B$ and orientation in the range of $B^r$; this is shown
in Fig.~\ref{fig:rod-Minkowski-sum} as the green ``square cone'' with
apex at $m_B$, four side planes (call them $H_1, \cdots H_4$) each of
which goes through $m_B$ and a side of $B^r$ (shown in blue), and the
green patch on the green sphere with radius $r_0$. Then we have
$\Fp(B) = \Fp(m_B, B^r) \oplus B^t$ (where $\oplus$ denotes the
Minkowski sum). First, we approximate it by a superset $\Fp(m_B, B^r)
\oplus D(r_B)$, where $D(r_B)$ is a ball with radius $r_B$. This
amounts to placing balls of radius $r_B$ centered at the 5 corners of
the green square cone (i.e., the 5 red balls $D_1 \cdots D_5$) and
taking the convex hull. We further approximate this convex hull by its
superset as follows. For each plane $H_i$ we move it from the green
square cone outward along its normal direction by a distance $r_B$;
call the resulting plane $H_i'$.  Note that $H_i'$ is parallel to $H_i$
and is tangent to 3 red balls. We consider the volume $V$ bounded by
the 4 planes $H_i'$ on the sides, on the top the pink sphere centered
at $m_B$ with radius $r_0 + r_B$, and at the bottom the plane $H_5'$
that is parallel to the plane of $B^r$ and tangent to $D_5$ at its
lowest point.
(Another natural choice to bound the bottom is to replace $H_5'$ by
the pink plane shown in Fig.~\ref{fig:rod-Minkowski-sum}, which is
tangent to $D_5$ with the plane normal being the axis of the square
cone. We choose $H_5'$ instead for a technical reason; see below.)
Clearly $V$ is a superset of the convex hull, and thus a superset of
$\Fp(B)$.  Now we define $\wtFp(B) = V$. Note that $\wtFp(B)$ is the
intersection of 5 half spaces bounded by $H_1', \cdots H_5'$, and the
pink ball (with center $m_B$ and radius $r_0 + r_B$). Checking if a
feature $f$ intersects $\wtFp(B)$ can be done by checking the half
spaces/ball one by one easily. (If $f$ is an edge or a wall, we need
to maintain the range/convex set of common intersections so far, and
at the end see if such range/convex set is non-empty or not).

	    	\begin{figure}[htb]
	    	  \begin{center}
	\includegraphics[scale=0.25]{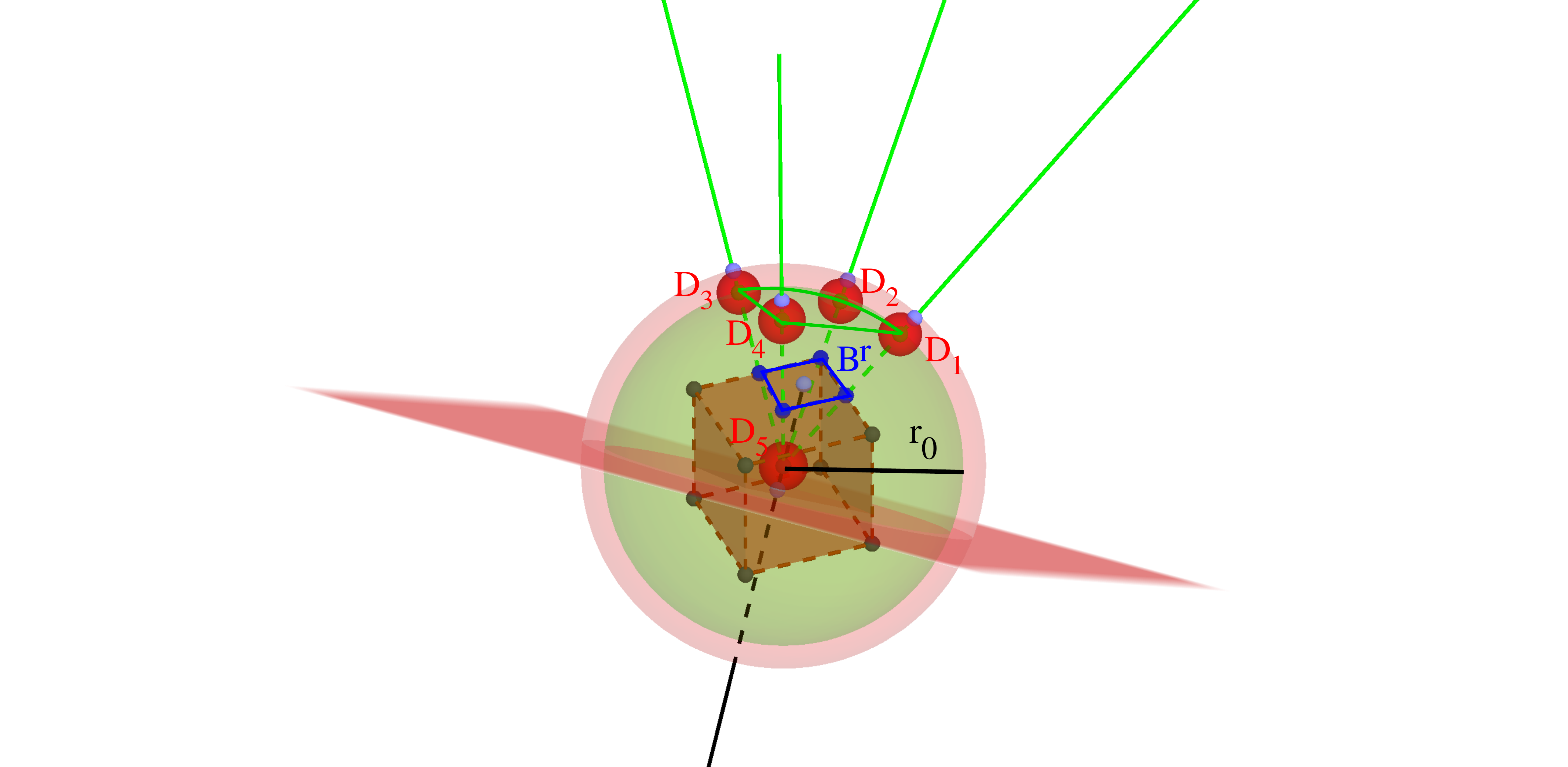}
	    	   \caption{The exact footprint of $B$, $\Fp(B)$, and its
                            superset $\wtFp(B)$.}
	    	   \label{fig:rod-Minkowski-sum}
	    	  \end{center}
	    	\end{figure} 
}

%
\noindent {\bf Theorem~3.} 
{\em
The approximate footprint $\wtFp(B)$ as defined for a rod robot
satisfies Eq.~\refeq{incl1}, i.e., there exists some fixed constant
$\sigma >1$ such that $\wtFp(B/\sigma) \ib \Fp(B) \ib \wtFp(B)$.
} 
\ \\ \noindent
\bpf
%
%
We have $\Fp(B) \ib \wtFp(B)$ by construction, so we just need to
prove that there exists some fixed constant $\sigma >1$ such that
$\wtFp(B/\sigma) \ib \Fp(B)$.
The idea is to first use a ``nice'' shape to contain $\wtFp(B)$, and
then show that we can shrink this nice shape by a factor of some fixed
constant $\sigma >1$ such that it is contained in $\Fp(B)$.
Let $c$ be the center of $B^r$.  Clearly the round cone $Cone_{round}
\as Cone(m_B, Ball(m_B + B^r)$ contains the square cone $Cone_{square}
\as Cone(m_B, B^r + m_B)$, and thus $V \as Cone_{round} \cap Ball(r_o,
m_B)$ contains $Cone_{square} \cap Ball(r_o, m_B) = Fp_0(B)$.
%
%
%
Recall that $\wtFp(B) = ``$\wtFp(B)$'' \cap H_0$.  Consider the point
$q$ on $H_0$ that is cut by ``$\wtFp(B)$'' and is farthest from
$m_B$. The distance between $q$ and $m_B$ depends on the orientation
of the square/round cone axis (going through $m_B$ and $c$). The
maximum happens when the axis goes from the center to the corner of
the brown box in Fig.~\ref{fig:rod-Minkowski-sum}, making an angle of
$\arcsin (1/\sqrt{3})$. Since the distance between $m_B$ and $H_0$ is
$r_B$, this maximum distance between $q$ and $m_B$ is $\sqrt{3}
r_B$. Therefore $\wtFp(B)$ is contained in $V_{final} \as V \oplus Ball(\sqrt{3}
r_B)$.
Also, $Cone_{round} / \sqrt{2}$ is contained in $Cone_{square}$. Note
that $Fp_0(B) \oplus (B^t - m_B) = \Fp(B)$, where $(B^t - m_B)$
contains $Ball(r_B / \sqrt{3})$. Now consider $V_{final} / 3$: $V /3$
is contained in $\Fp_0(B)$ and $Ball(\sqrt{3} r_B) / 3 = Ball(r_B /
\sqrt{3})$ is contained in $(B^t - m_B)$, and thus $V_{final} / 3 \ib
\Fp(B)$. Overall, we have $\wtFp(B/3)  \ib V_{final} / 3 \ib \Fp(B)$.
\epf

Note that the existence of such a constant $\sigma$ is all we need
to guarantee that our algorithm is resolution-exact; we do not need
to know this constant in implementations.

\remove{
%
\noindent {\bf Lemma~2.} 
{\em (Inheritance) \\
Let $B_1$ be a child of $B$. Then $\wtFp(B_1) \ib \wtFp(B)$.
(This means that we can obtain $\wtphi(B_1)$ by distributing the
features from the parent feature set $\wtphi(B)$ to the children.)
} \\
\ \\ \noindent
\bpf
Let $B= B^t \times B^r$ and $B_1 = B_1^t \times B_1^r$.  We consider
two cases: (1) $B_1^t$ is a child of $B^t$ but their rotational boxes
are the same, and (2) $B_1^r$ is a child of $B^r$ but their
translational boxes are the same.
For (1), their inner footprints $\Fp_0(B)$ and $\Fp_0(B_1)$ (each a
green square cone intersected by the green ball in
Fig.~\ref{fig:rod-Minkowski-sum}) are of the same shape and
orientation; only the positions are shifted: each has the apex at its
own translational box center. The apices (so are all other
corresponding corners where red balls $D_i$ in
Fig.~\ref{fig:rod-Minkowski-sum} are centered) are apart by a distance
of $r_{B_1}$; the red balls $D_i$ of $B_1$ are of radius $r_{B_1}$,
but for the parent $B$ the red-ball radius is twice as large ($r_B = 2
r_{B_1}$). Thus each red ball of the child is entirely contained in
the corresponding red ball of the parent, and thus $\wtFp(B_1) \ib
\wtFp(B)$.

For (2), now their inner footprints have the same apex at $m_B$, and
their red balls $D_i$ have the same radius $r_B$.  Since the child
angular range $B_1^r$ is a subset of the parent angular range, the
green square cone (Fig.~\ref{fig:rod-Minkowski-sum}) of the child is
entirely contained in the green square cone of the parent. Thus for
``$\wtFp(B)$'', i.e., before bounding by $H_0$ to get $\wtFp(B)$ 
(recall that $\wtFp(B) \as ``\wtFp(B)\textrm{''} \cap H_0$; see also
\refeq{sqcone2}), the desired containment property holds.
%
%
The bottom part bounded by $H_0$ has no problem either, since their
horizontal planes $H_0$ are the same.
(Note that if we replace $H_0$ by the pink plane in
Fig.~\ref{fig:rod-Minkowski-sum} then it will be tilted differently
since the square cone axis is oriented differently for $B_1$ and $B$.)
\epf
\ \\

\noindent {\bf Lemma~3.}
{\em
There exists some fixed constant $\sigma >1$ such that $\wtFp(B /
\sigma) \ib \Fp(B)$.
} \\
\ \\ \noindent
\bpf
The idea is to first use a ``nice'' shape to contain $\wtFp(B)$, and
then show that we can shrink this nice shape by a factor of some fixed
constant $\sigma >1$ such that it is contained in $\Fp(B)$.
Let $c$ be the center of $B^r$.  Clearly the round cone $Cone_{round}
\as Cone(m_B, Ball(m_B + B^r)$ contains the square cone $Cone_{square}
\as Cone(m_B, B^r + m_B)$, and thus $V \as Cone_{round} \cap Ball(r_o,
m_B)$ contains $Cone_{square} \cap Ball(r_o, m_B) = Fp_0(B)$.
%
%
%
Recall that $\wtFp(B) = ``$\wtFp(B)$'' \cap H_0$.  Consider the point
$q$ on $H_0$ that is cut by ``$\wtFp(B)$'' and is farthest from
$m_B$. The distance between $q$ and $m_B$ depends on the orientation
of the square/round cone axis (going through $m_B$ and $c$). The
maximum happens when the axis goes from the center to the corner of
the brown box in Fig.~\ref{fig:rod-Minkowski-sum}, making an angle of
$arcsin (1/\sqrt{3})$. Since the distance between $m_B$ and $H_0$ is
$r_B$, this maximum distance between $q$ and $m_B$ is $\sqrt{3}
r_B$. Therefore $\wtFp(B)$ is contained in $V_{final} \as V \oplus Ball(\sqrt{3}
r_B)$.
Also, $Cone_{round} / \sqrt{2}$ is contained in $Cone_{square}$. Note
that $Fp_0(B) \oplus (B^t - m_B) = \Fp(B)$, where $(B^t - m_B)$
contains $Ball(r_B / \sqrt{3})$. Now consider $V_{final} / 3$: $V /3$
is contained in $\Fp_0(B)$ and $Ball(\sqrt{3} r_B) / 3 = Ball(r_B /
\sqrt{3})$ is contained in $(B^t - m_B)$, and thus $V_{final} / 3 \ib
\Fp(B)$. Overall, we have $\wtFp(B/3) /  \ib V_{final} / 3 \ib \Fp(B)$.
\epf

Note that the existence of such a constant $\sigma$ is all we need
to guarantee that our algorithm is resolution-exact; we do not need
to know this constant in implementations.

} 

\section{Appendix: Soft Predicate for a Ring -- Proofs}
%
\subsection{Computing the Separation Between a Circle and a Feature} \label{appendixE-Separation}
As mentioned in Sec.~\ref{se-soft-ring}, our soft predicates for the
ring robot need to compute the separation of an embedded circle $C$
from $f$, i.e., $\Sep(C,f)$, where $f$ is a point, line or a plane.

In the following, let $C$ be a circle of radius $r$
centered at $O$, and lying in a plane $P_C$ with normal vector $n$.
Also let $u$ be a vector along the direction of line $L$.
Note that $r,n,O,u$ are all given constants.

\medskip \noindent {\bf Simple Filtering}\\
Before actually computing $\Sep(C,f)$, we can first perform a simple
filtering.  Recall from Sec.~\ref{se-soft-ring} that the purpose of
$\Sep(C,f)$ is to decide ``Is $\Sep(C, f) \le r_B$?''. If we have a
simple way to know that $\Sep(C, f) > r_B$ then there is no need to
compute $\Sep(C,f)$. Here is how. Suppose $f$ is a line or a plane.
We can easily compute the separation $d$ from the circle center $O$ to
$f$, i.e., $d = \Sep(O, f)$. If $d > r + r_B$, then $\Sep(C, f) \ge d
- r > r_B$ and we are done. Only when $d \le r + r_B$ do we need to
compute $\Sep(C,f)$, which can be much more complicated (see below).

\medskip \noindent {\bf Computing the Separation $\Sep(C,f)$}\\
The case where $f$ is a point is trivial, and involves solving a
quadratic equation.  The case $f$ is a plane is a rational problem: if
$f$ is parallel to $P_C$, then $\Sep(C,f)$ is just the separation
between the two planes.  Otherwise, let $L'$ be the intersection of
the two planes.  Let $p\in C$ be the closest point in $C$ to $L'$, and
$q$ the projection of $p$ to the plane $f$.  Then $\Sep(C,f)=\|p-q\|$.
(Note: if $L'$ intersects $C$, then $p$ is just any point in $L'\cap
C$ and $p=q$ in this case.)

Finally, we address the most interesting case, where $f$ is a line $L$
defined by an obstacle edge.  
But before showing the exact computation of $\Sep(C,L)$, we
show a relatively easy way to compute an upper bound, denoted
$\Sep'(C,L)$, on $\Sep(C,L)$.
	We project the two edge endpoints $p_1, p_2$ onto
	the plane $P_C$ to get $p_1', p_2'$.
	First, assume $p_1'\neq p_2'$ (non-degenerate case).
	Then any point in this projected line $L'$
	is expressed by $p_1' + t (p_2' - p_1')$ with parameter $t$. Let $p'$
	be the point in $L'$ closest to $C$; recall that $O$ is the circle
	center. 
	The corresponding point $p \in L$ that projects to $p'$
	has the same $t$ as $p$.  Then
	we compute $\Sep(C, p') \as d$ from the radius and the
	distance between $p'$ and $O$. Suppose $q$ is the point on $C$ closest
	to $p'$. Then define $\Sep'(C, L) \as ||p - q||$.
	We can obtain $||p - q||$
	without solving $q$, by the fact that $q, p', p$ form a right triangle
	with leg lengths $d$ and $||p - p'||$.
	We return to the degenerate case where
	$p_1'=p_2'$.   This means 
	$L$ is perpendicular to $P_C$, and $\Sep(C,L)$
	is easily obtained.  But numerically, whenever $\|p_1'-p_2'\|$
	is small, we ought to use this particular approximation.
	Since this is just a filter, we will not dwell on this.

\remove{ 
	CHECK... \\
        YJC: The above assumes that $L'$ is outside $C$. But $L'$ may
        intersect $C$... \\
} 

\medskip \noindent {\bf Reduction of $\Sep(C,f)$ to Root-Finding}\\
	We now show how to reduce computing $\Sep(C,L)$ to solving quartic
	equations.
	Let $p, q$ be the two points with $p\in
	C$ and $q\in L$ such that $\Sep(C,L)=\|p-q\|$.  We can view
	$p=p(x,y,z)$ and $q=q(t)$ where $x,y,z,t$ are variables to be solved.
	
	We obtain four equations by the following conditions.
	    \begin{description}

    \item[(A)] The point $p$ lies in the sphere centered at $O$ of radius $r$:
		\beql{pr}
		\|p-O\|=r^2.
		\eeql
		Explicitly, $(x-O_x)^2+ (y-O_y)^2+ (z-O_z)^2 = r^2$.
    \item[(B)] The plane $Opq$ is perpendicular to the plane of $C$:
		\beql{Opq}
		((p-O)\times (q-O))\cdot n = 0.
		\eeql
		This equation is multilinear in $t$ and in $\set{x,y,z}$.
		It has the form $t A(x,y,z)+ B(x,y,z,t) + C=0$
		where $A,B$ are linear in the indicated variables,
		and $C$ is a constant.
    \item[(C)] The line $pq$ is perpendicular to $L$:
		\beql{pqu}
		(p-q)\cdot u = 0.
		\eeql
		This is a linear function in $x,y,z,t$.
    \item[(D)] The (radius) line $Op$ is perpendicular to $n$:
		\beql{pn}
		(p-O)\cdot n = 0.
		\eeql
		This is a linear function in $x,y,z$.
    \end{description}

	Using Condition~(D), 
        we can express $z$ as a linear function in $x,y$
	and plug into Eqs.\ of~(A), (B), (C)  
        to eliminate $z$ without
	changing the nature of these equations (i.e., Eq.\ of (A) 
        remains quadratic and Eq.\ of (B) 
        remains multilinear).
	By using Condition~(C) 
        we can eliminate $t$ from Eq.\ of (B) 
        and turn it  
        into a quadratic equation in $x,y$.  
        So we now have a system of two quadratic equations in $x,y$:
		\beql{biquadratic}
		\grouping{ax^2 + bx +c &=& 0\\
		a'x^2 + b'x +c' &=& 0}
		\eeql
	where $a,b,c$ (resp., $a',b',c'$) are polynomials in $y$
	of degrees $0,1,2$ respectively. We obtain
		$ x = \frac{-b \pm\sqrt{\Delta}}{2a}
		  = \frac{-b' \pm\sqrt{\Delta'}}{2a'}$
	where $\Delta=b^2-4ac$ and $\Delta'$ similarly.
	Thus
		\beqarrys
		a'(-b\pm\sqrt{\Delta})
			&=& a(-b'\pm\sqrt{\Delta'})\\
		A \pm a'\sqrt{\Delta}
			&=& \pm a\sqrt{\Delta'} & \wherE\ A=\det\mmat{a&b\\a'&b'}\\
		\Big(A \pm a'\sqrt{\Delta}\Big)^2
			&=& a^2\Delta'\\
		\pm 2a'A\sqrt{\Delta}
			&=& a^2\Delta' - A^2- (a')^2\Delta\\
		(2a'A)^2 \Delta
			&=& \Big(a^2\Delta' - A^2 - (a')^2\Delta\Big)^2.
		\eeqarrys
          We summarize by restating the last equation:
       
		\beql{quartic}
		(2a'A)^2 \Delta
			= \Big(a^2\Delta' - A^2 - (a')^2\Delta\Big)^2.
		\eeql
	This is a quartic equation in $y$, as claimed.

\subsection{Proof of Properties} \label{appendixE-Proof}
%
%
\noindent {\bf Theorem~4.} 
%
%
{\em
The approximate footprint $\wtFp(B)$ as defined for a ring robot
satisfies Eq.~\refeq{incl1}, i.e., there exists some fixed constant
$\sigma >1$ such that $\wtFp(B/\sigma) \ib \Fp(B) \ib \wtFp(B)$.
} 
%
%
\ \\ \noindent
\bpf
%
%
%
%
We have $\Fp(B) \ib \wtFp(B)$ by construction, so we just need to
prove that there exists some fixed constant $\sigma >1$ such that
$\wtFp(B/\sigma) \ib \Fp(B)$.
Recall that $\wtFp(B) = \Fp_1(B) \oplus Ball(r_B)$.  For $\Fp(B)$, it
is the Minkowski sum of $Fp_0(B)$ and a cube of radius $r_B$. The
difference between $Fp_0(B)$ and $\Fp_1(B)$ is the orientation of the
cone axis, with the maximum difference happening when the axis goes
from the cube center to a cube corner, making a factor of $\sqrt{3}$.
For the other part of the Minkowski sum, $Ball(r_B / \sqrt{3})$ is
contained in a cube of radius $r_B$. Overall, the statement is true
with $\sigma = \sqrt{3}$.
\epf

%
\remove{ 
%
%
\noindent
{\bf Theorem~1.} 
{\em
$\wtFp(B)$ satisfies the inclusion properties in Eqs.~\refeq{incl1}
  and \refeq{incl2}. Namely:\\
(a) (Inheritance)
Let $B_1$ be a child of $B$. Then $\wtFp(B_1) \ib \wtFp(B)$.\\
(b) There exists some fixed constant $\sigma >1$ such that $\wtFp(B /
\sigma) \ib \Fp(B)$.
} \\
\bpf
\\
{\bf Part (a).}
We would like to have the feature-set inheritance property from parent
to children. But it seems not easy to prove it directly. Instead, we
{\em enforce} this property {\em algorithmically}. For clarity, let
us re-name $\wtFp(B)$ defined in Sec.~\ref{se-soft-ring} as
$\wtFp'(B)$.  
%
%
  As given,
$B_1$ is a child of $B$.  In computing the child feature set, we
  always take the parent feature set and test against $\wtFp'(B_1)$
  for the child $B_1$. This is equivalent to setting $\wtFp(B_1) =
  \wtFp'(B_1) \cap \wtFp(B)$. The question is whether $\wtFp(B_1)$ is
  still a superset of $\Fp(B_1)$ as required in Eq.~\refeq{incl1}.

\noindent {\bf Claim:}
$\Fp(B_1) \ib  \wtFp(B_1)$. 
%
\\
{\bf Proof of Claim:}
%
By construction $\wtFp'(B_1)$ is a superset of $\Fp(B_1)$. It suffices
to show that $\wtFp(B)$ is a superset of $\Fp(B_1)$. But $\wtFp(B)$ is
a superset of $\Fp(B)$ (initially and inductively), which in term is a
superset of $\Fp(B_1)$.
This completes the proof of Claim. $\Box$ \\

\noindent
{\bf Part (b).}
%
%
  When
trying to prove $\wtFp(B / \sigma) \ib
\Fp(B)$ for some fixed constant $\sigma >1$, we can just use
$\wtFp'(B)$ and try to prove $\wtFp'(B / \sigma) \ib \Fp(B)$ without
worrying about its intersection with $\wtFp(\mbox{parent})$ since the
intersection is a smaller set.
\remove{ 
\blem 
There exists some fixed constant $\sigma >1$ such that $\wtFp(B /
\sigma) \ib \Fp(B)$.
\elem
\bpf
As said above, 
} 
Therefore,
we only need to consider $\wtFp'(B) = \Fp_1(B) \oplus
Ball(r_B)$.  For $\Fp(B)$, it is the Minkowski sum of $Fp_0(B)$ and
a cube of radius $r_B$. The difference between $Fp_0(B)$ and
$\Fp_1(B)$ is the orientation of the cone axis, with the maximum
difference happening when the axis goes from the cube center to a cube
corner, making a factor of $\sqrt{3}$.  For the other part of the
Minkowski sum, $Ball(r_B / \sqrt{3})$ is contained in a cube of
radius $r_B$. Overall, the statement is true with $\sigma = \sqrt{3}$.
\epf
%

%
%
\remove{ 
We would like to have the feature-set inheritance property from parent
to children. But it seems not easy to prove it directly. Instead, we
{\em enforce} this property {\em algorithmically}. For clarity, let
us re-name $\wtFp(B)$ defined in Sec.~\ref{se-soft-ring} as
$\wtFp'(B)$.  Suppose $B_1$ is a child of $B$.  In computing the child
feature set, we always take the parent feature set and test against
$\wtFp'(B_1)$ for the child $B_1$. This is equivalent to setting
$\wtFp(B_1) = \wtFp'(B_1) \cap \wtFp(B)$. The question is whether
$\wtFp(B_1)$ is still a superset of $\Fp(B_1)$ as required in Eq.~\refeq{incl1}.
\blem
$\Fp(B_1) \ib  \wtFp(B_1)$. 
\elem
\bpf
By construction $\wtFp'(B_1)$ is a superset of $\Fp(B_1)$. It suffices
to show that $\wtFp(B)$ is a superset of $\Fp(B_1)$. But $\wtFp(B)$ is
a superset of $\Fp(B)$ (initially and inductively), which in term is a
superset of $\Fp(B_1)$.
\epf

For the following lemma, when trying to prove $\wtFp(B / \sigma) \ib
\Fp(B)$ for some fixed constant $\sigma >1$, we can just use
$\wtFp'(B)$ and try to prove $\wtFp'(B / \sigma) \ib \Fp(B)$ without
worrying about its intersection with $\wtFp(\mbox{parent})$ since the
intersection is a smaller set.

\blem 
There exists some fixed constant $\sigma >1$ such that $\wtFp(B /
\sigma) \ib \Fp(B)$.
\elem
\bpf
As said above, we only need to consider $\wtFp'(B) = \Fp_1(B) \oplus
Ball(r_B)$.  For $\Fp(B)$, it is the Minkowski sum of $Fp_0(B)$ and
a cube of radius $r_B$. The difference between $Fp_0(B)$ and
$\Fp_1(B)$ is the orientation of the cone axis, with the maximum
difference happening when the axis goes from the cube center to a cube
corner, making a factor of $\sqrt{3}$.  For the other part of the
Minkowski sum, $Ball(r_B / \sqrt{3})$ is contained in a cube of
radius $r_B$. Overall, the statement is true with $\sigma = \sqrt{3}$.
\epf
} 
%
} 
%
%

\ignore{ 
Now we consider the robot $R_0$ to be a ring, which is a circle with
center $A$ and radius $r_0$. We specify its configuration by the
position of $A$ and the orientation of its axis, where the axis goes
through $A$ and is perpendicular to the plane that the circle lies on.
As before, let $B = B^t \times B^r$ be a subdivision box. Our main
task is to consider $\Fp(B)$, the (exact) footprint of $R_0$ when
moving around inside the box $B$, and construct its desirable superset
$\wtFp(B)$, for which we can easily compute whether a feature $f$
intersects $\wtFp(B)$.

We start by considering $\Fp(m_B, B^r)$, the (exact) footprint of the
ring with $A$ at $m_B$ and the orientation of the axis in the range of
$B^r$. Let $d$ be the center point of the square $B^r$. Note that
$B^r$ is not very easy to work with directly, since the ring axis
going through $d$ is tilted, i.e., not perpendicular to the plane of
$B^r$. Rather, we work on a (round) cone with apex at $m_B$ and cone
axis going through $d$, such that the cone just contains $B^r$ (and we
already approximate $B^r$ by a superset; the cone is defined by the
apex $m_B$ and the ball around $d$ with radius half the diameter of
$B^r$ such that the cone just contains that ball). Now the orientation
of the ring axis is specified by the angle $\theta$ relative to the
cone axis. In this way, the footprint is symmetric.

Now we consider the footprint $\Fp(m_B, [0,\Theta])$, where
$[0,\Theta]$ is the range of the angle $\theta$ and $\Theta$ is the
maximum angle the cone allows ($\Fp(m_B, [0,\Theta])$ is a superset of
$\Fp(m_B, B^r)$ since some $\theta$ value corresponds to an angle
outside $B^r$). It is easier to look at the cross-section plane that
contains the cone axis and the ring axis. The ring shows up on this
plane at two points; as $\theta$ changes over the range $[0,\Theta]$,
$\Fp(m_B, [0,\Theta])$ becomes two arcs on this plane that are parts
of a circle centered at $m_B$ with radius $r_0$. Since $\Fp(B) =
\Fp(m_B, B^r) \oplus B^t$, we can approximate it by $\Fp(m_B,
   [0,\Theta]) \oplus D(r_B)$ where $D(r_B)$ is a ball of radius
   $r_B$. This amounts to putting a circle of radius $r_B$ around the
   4 endpoints of the circular arcs on the cross-section plane to get
   the 4 red circular arcs each tangent to two of the 4 circles we
   put, and rotating the cross-section plane around the cone
   axis. 
%
%
   Ideally, we want to take $\Fp(m_B,
   [0,\Theta]) \oplus D(r_B)$ as $\wtFp(B)$, but not yet;
   we call it $\wtFp'(B)$ instead (see technical reasons below).
   It is the {\em union} of the following
   3 parts:\\
(1) the middle part, the sliced annulus $SA$, which is the
   intersection of the outer ball centered at $m_B$ with radius $r_0 +
   r_B$, the complement of the inner ball centered at $m_B$ with
   radius $r_0 - r_B$, sliced by intersecting the two half spaces
   defined by two parallel planes $P_1$ and $P_2$ ($P_1$ is above
   $P_2$ and $m_B$ is in between, with the same distance $r_0 \sin
   \Theta$ to $P_i$) that cut through the outer and inner balls; \\
(2) the top part, which is $C_1 \oplus D(r_B)$,
    the Minkowski sum of a circle $C_1$ and a
   ball $D(r_B)$, where $C_1$ lies on $P_1$, centered at the
   intersection of the cone axis and $P_1$, with radius $r_0 \cos
   \Theta$;\\
(3) the bottom part, $C_2 \oplus D(r_B)$, similar to (2); we call the
   circle $C_2$.

Let $B_1$ be a child of $B$. As in Lemma~\ref{le-rod-inheritance}, we
  would like to show that $\wtFp'(B_1) \ib \wtFp'(B)$, so that the
  soft version of the feature set can be distributed from $B$ to
  $B_1$. However, it seems not easy to show $\wtFp'(B_1) \ib
  \wtFp'(B)$ as their cones are tilted differently. Instead, we {\em
    force} the inheritance/distribution property of the feature set
        {\em algorithmically}: in computing the child feature set, we
        always take the parent feature set and test against
        $\wtFp'(B_1)$ for the child $B_1$. This is equivalent to
        setting $\wtFp(B_1) = \wtFp'(B_1) \cap \wtFp(B)$. The question is
whether $\wtFp(B_1)$ is still a superset of $\Fp(B_1)$.

\blem
$\Fp(B_1) \ib  \wtFp(B_1)$. 
\elem
\bpf
By construction $\wtFp'(B_1)$ is a superset of $\Fp(B_1)$. It suffices
to show that $\wtFp(B)$ is a superset of $\Fp(B_1)$. But $\wtFp(B)$ is
a superset of $\Fp(B)$ (initially and inductively), which in term is a
superset of $\Fp(B_1)$.
\epf

\blem 
There exists some fixed constant $\sigma >1$ such that $\wtFp(B /
\sigma) \ib \Fp(B)$.
\elem
%
%
%
%
} 

\ignore{
%
\subsection{Intersection Tests for a Ring}
To test whether a feature $f$ intersects $\wtFp'(B)$, recall that
$\wtFp'(B)$ has 3 parts: (1) the middle part, the sliced annulus $SA$,
(2) the top part $C_1 \oplus D(r_B)$, and (3) the bottom part $C_2
\oplus D(r_B)$.  (1) is the intersection of two half spaces, one ball,
and the complement of a ball, and we can test the intersection as
usual. For (2) and (3), we check if $\Sep(f,C_i) \leq r_B$.

We use $C$ to denote the circle $C_i$, and $P_C$ the plane that $C$
lies on. Let $O = (x_0, y_0, z_0)$ be the center of $C$. Note that $C$
is the intersection of the plane $P_C$ and the sphere $S_C$ centered
at $O$ with radius $r_0 \cos \Theta$. We consider three cases in that
order: (A) $f$ is a corner (point); (B) $f$ is a line segment; (C) $f$
is a triangle.

%
\ \\
{\bf Case (A):} $f$ is a corner, i.e., a point $p$. To compute
$\Sep(f,C) = \Sep(C, p)$, we want to find the point $q \in C$ such
that $q$ and $p$ are the closest, so that $||p - q|| = \Sep(C, p)$. We
can find $q$ as follows: Project $p$ onto the plane $P_C$ at point
$p'$, make a line going through $p'$ and $O$ to intersect $C$ at two
points $q, q'$; the one closer to $p'$ is the desired $q$.
%
%
Computationally, let $(x_q, y_q, z_q)$ be the coordinates of $q$. The
fact that $O, q, p'$ are co-linear gives two linear equations: 
\begin{quote}
(a) $\frac{x_q - x_0}{x_{p'} - x_0} = \frac{y_q - y_0}{y_{p'} - y_0}$,
and \\
(b) $\frac{x_q - x_0}{x_{p'} - x_0} = \frac{z_q - z_0}{z_{p'} - z_0}$, 
\end{quote}
where $p' = (x_{p'}, y_{p'}, z_{p'})$ is known.  From (a) we can
easily express $y_q$ in terms of $x_q$ linearly, and similarly for
$z_q$ from (b). Substituting them into the equation that $q$ is on the
sphere $S_C$, we get a quadratic equation of a single variable $x_q$,
which can be easily solved to obtain two roots for $q$.
(Note that there is another linear equation, i.e., (b') $q = (x_q,
y_q, z_q)$ satisfies the plane equation of $P_C$. But (a) and (b)
together means that $q$ is linear between $O$ and $p'$, and since
$O$ and $p'$ are both on $P_C$, the plane equation in (b') is
automatically consistent with (a) and (b).)

%
\ \\
{\bf Case (B):} $f$ is a line segment with endpoints $p_1, p_2$. We do
this test only if $\Sep(C, p_1) > r_B$ and $\Sep(C, p_2) > r_B$ by the
tests of {\bf Case (A)} (otherwise $f$ is already included to the
feature set $\wtphi_i(B)$). Let $L$ be the line going through $p_1$
and $p_2$.  Our main task is to find a point $p \in L$ and a point $q
\in C$ such that $||p - q|| = \Sep(C,L)$. We can write $p$ as $p = p_1
+ t (p_2 - p_1)$ where $t$ is an unknown variable. 
If the resulting $p$ does not lie between $p_1$ and $p_2$ (i.e., $t$
is outside the range $(0,1)$), then we can immediately stop the test
and decide that $f$ does not belong to the feature set, {\em without
  even solving $q$!} (This is because the closest point of $f$ to $C$
will be either $p_1$ or $p_2$, but we already have $\Sep(C, p_i) >
r_B$ for $i = 1, 2$ by the tests of {\bf Case (A)}).
(As we will see later, the best way is to solve $p$ first, which is
easy (a linear equation of one variable), and then use $p$ to solve
$q$ (which amounts to a quadratic equation of one variable).  Checking
$t$ first may result in an early termination to avoid computing $q$.)
Otherwise, $p$ is between $p_1$ and $p_2$ (i.e., $t \in (0,1)$). Then
$f$ belongs to the feature set iff $||p - q|| (= \Sep(C,L)) \leq r_B$.

Now the task is to find $p \in L$ and $q \in C$ such that $||p - q|| =
\Sep(C,L)$. Recall that $p = p_1 + t (p_2 - p_1)$ for a variable
$t$. The condition that $p$ and $q$ define $\Sep(C,L)$ means that the
vector $p - q$ is perpendicular to both the line $L$ and a tangent
line to $C$ going through $q$. We project $p_1, p_2$ onto the plane
$P_C$ to get points $p_1', p_2'$; call $L'$ and $p'$ the projections
of $L$ and $p$ onto $P_C$. Then we have $p' = p_1' + t (p_2' - p_1')$
for the same $t$. Also, on the same plane $P_C$, the vector $p' - q$
is perpendicular to both $L'$ and the tangent to $C$ at $q$. Note that
the latter condition means that {\em $O, q, p'$ are co-linear}
since the vector $q - O$ is also perpendicular to the same tangent at $q$.
Then $p' - q$ and $p' - O$ are parallel, and we have 
\begin{quote}
    (a) $\bang{p' - O, p_2' - p_1'} = 0$ (i.e., $p' - O$ is perpendicular to $L'$). 
\end{quote}
Note that (a) is a linear equation with only one variable $t$, which
we can solve easily. At this point, if $t$ is outside the range
$(0,1)$ then we can already decide that $f$ does not belong to the
feature set and stop the test (see the discussions above). Otherwise
($t \in (0,1)$), we go ahead and use $t$ to obtain $p'$ and $p$. Now we
have the conditions that $O, q, p'$ are co-linear and that $q$ is on
$C$, for 3 variables $(x_q, y_q, z_q) = q$ --- this is exactly the
same as in {\bf Case (A)} (where co-linearity gives two equations (a)
and (b) of {\bf Case (A)}, plus the equation that $q$ is on the sphere
$S_C$). Again, this amounts to solving a quadratic equation of one
variable, and between the two roots we can choose the correct one for
$q$ in the same way.
Finally, from $||p - q|| = \Sep(C,L)$, $f$ belongs to the feature set
iff $||p - q||  \leq r_B$.

%
\ \\
{\bf Case (C):} $f$ is a triangle on a plane $P_T$.  We do this test
only if for {\em each} boundary element $\partial f$ (a vertex or an edge) of
$f$ we have $\Sep(C, \partial f) > r_B$ by the tests in {\bf Cases (A)}
and {\bf (B)} (otherwise $f$ is already included to the feature set).
Now the task is to find a point $p \in P_T$ and a point $q \in C$ such
that $||p - q|| = \Sep(C, P_T)$. 
%
%
%
If $||p - q|| > r_B$ then $f$ does not belong to the feature
set. Otherwise ($||p - q|| \leq r_B$) then $f$ belongs to the feature
set iff $p$ is in the interior of the triangle. (Note: If $p$ is
outside the triangle, then the closest point on $f$ to $C$ must lie on
some boundary element $\partial f$, which is already known to have
$\Sep(C, \partial f) > r_B$, and thus $f$ does not belong to the
feature set.) 
(To check whether a point $p \in P_T$ is inside the triangle, let the
triangle vertices be $p_1, p_2, p_3$; any point inside the triangle
satisfies $(x, y, z) = p_1 + u (p_2 - p_1) + v (p_3 - p_1)$, where $u,
v \in [0,1]$ and $u + v \leq 1$. We can use the coordinates of $p$ to
solve $u$ and $v$ and check if their constraints are satisfied.)
%
%

To find a point $p \in P_T$ and a point $q \in C$ such that $||p - q||
= \Sep(C, P_T)$, we do the following. Let $L'$ be the line intersected
by the two planes $P_T$ and $P_C$, and $n_T$ and $n_C$ the normal
vectors of $P_T$ and $P_C$. The direction of $L'$ is given by the
vector $v_{L'} = n_T \times n_C$ where $\times$ is the cross
product. We can find a point $p_0 \in L'$ by letting one of $x, y, z$
be 0 and solve the system of two linear equations given by the plane
equations of $P_T$ and $P_C$. Suppose $p' \in L'$ and $q \in C$
define $\Sep(C, L')$, i.e., $||p' - q|| = \Sep(C, L')$. We can
express $p'$ as $p' = p_0 + t \cdot v_{L'}$ with one variable $t$. Again, on
the same plane $P_C$, $p' - q$ is perpendicular to both the line $L'$
and the tangent to $C$ at $q$. As discussed in {\bf Case (B)}, this
means that $O, q, p'$ are co-linear and we have 
\begin{quote}
(a) $\bang{p' - O, v_{L'} } = 0$ (i.e., $p' - O$ is perpendicular to $L'$), 
\end{quote}
which is a linear equation on $t$ and we can solve $t$ (and $p'$)
easily. Now, with $p'$ known and the condition that $O, q, p'$ are
co-linear, we can solve $q$ with a quadratic equation of one variable
and choose the correct root (see {\bf Case (A)}). Finally, we project
$q$ back onto the plane $P_T$ to obtain the desired point $p$ such
that $||p - q|| = \Sep(C, P_T)$.
} 

\section{Appendix: Correct Implementation of Soft Exact Algorithms} \label{app-implement}
%
%
%
%
    The
    earlier sections provide
%
        an ``exact'' description of planners
	for a rod and a ring, albeit a ``soft kind''
	that admits a user-controlled amount of numerical indeterminacy.
	The reader may have noticed
	that we formulated precise mathematical relations
	and exact geometric shapes for which various inclusions
	must be verified for correctness.   Purely numerical
	computations (even with arbitrary precision)
	cannot ``exactly determine'' such relations in general.
	Nevertheless,
	we claim that all our computations can be guaranteed
	in the soft sense.  The basic idea is that for
	each box $B$, all the computations
	associated with $B$ is computed to some absolute
	error bound that at most $r_B/K^*$ where $r_B$
	is the box radius and $K^*$ is a constant depending on the
	algorithm only.  Thus, as
	boxes become smaller, we need higher precision
	(but the resolution $\vareps$ ensures termination).
	Moreover, the needed precision requires no special programming
	effort.  
	
	This is possible because all the inequalities in our
	algorithms are ``one-sided'' in the sense that
	we do not assume that the failure of an inequality test implies
	the complementary condition (as in exact (unqualified) computation).
	We can define a
	\dt{weak feature set} denoted $\whphi(B)$ with this property:
	\[ \whphi(B/\sigma)\ib \wtphi(B)\ib \whphi(B) \]
	for some $\sigma>1$.
	The ``weak'' $\whphi(B)$ is
	not uniquely determined (i.e., $\whphi(B)$ can be {\em any} set
	that satisfies the inequalities).  
	In contrast, the set $\wtphi(B)$ is mathematically precise and
	unique.  If we use $\whphi(B)$ instead of $\wtphi(B)$,
	the correctness of our planner remains intact. 
	Moreover, the weak set
	$\whphi(B)$  can be achieved as using numerical approximation
	(note: we do not need "correct rounding" from our bigFloats,
	so \gmp\ suffice). 

	We stress that these ideas have not been implemented, partly
	because there is no pressing need for this at present.

\section{Appendix: Counterexample for the Ring Heuristic}

We show that the use of $\Sep'(C,f)$
(Appendix~\ref{appendixE-Separation}) 
can lead to a wrong classification of a box $B$.  Recall that
$\Sep'(C,f)$ is an upper bound on $\Sep(C,f)$, and is an equality in
case $f$ is a corner or a triangle.

Assume that the footprint of configuration $m_B$ is
a unit circle $C$ centered at the origin lying in the horizontal
$z=0$ plane.  

We consider the polyhedral set $F\ib\RR^3$ such that the
intersection of $F$ with any horizontal plane $H:\set{z=z_0}$ (for any $z_0$)
is the L-shape $[-10,10]^2 \setminus (2,10]^2$ when projected to the
$(x,y)$-plane.
See \refFig{ring-robot-counter-example}.

\FigEPS{ring-robot-counter-example}{0.4}{Counterexample.} 

Let $f_0$ be the boundary feature of $F$ that is closest to circle $C$.
Clearly, $f_0$ is the vertical line $\bang{x=2,y=2}$.
Moreover, $\Sep(C,f_0)=2\sqrt{2}-1< 1.82$.
Now, slightly perturb $F$ so that $f_0$ is
slightly non-vertical, but it's projection onto the $(x,y)$-plane
is the line $y=2$
(in \refFig{ring-robot-counter-example},
$f_0$ is the red dot, and $y=2$ is the green line).
We also verify that $\Sep'(C,f_0)=\sqrt{5} \simeq 2.36$.

It is also important to see that
all the other boundary features $f\neq f_0$ of $F$, we
have $\Sep'(C,f)> 2$.  To see this, there are 2 possibilities for $f$:
if $f$ is an edge, this is clear.  If $f$ is a face, this is also clear unless
the face is bounded by $f_0$ (there are two such faces).  In this
case, our algorithm sets $\Sep'(C,f)$ to $\Sep'(C,f_0)$ which is $>2.23$.
Note that $F$ does not have any corner features.

Now construct any convex polyhedron $G\ib\RR^3$ that is disjoint
from $F$ such that boundary feature of $G$ that is closest to $C$
is a corner  $g_0 = (2.1, 2.1, 0)$.
It is easy to construct such a $G$.
Moreover, we see that $\Sep(C,g_0)=\Sep'(C,g_0)= \sqrt{2(2.1)^2} -1 \simeq
1.97$.

Suppose $\Omega=F\cup G$ and the translational and rotational parts of $B$
are given by $B^t=[-1/2,1/2]^2$
and $B^r=[-1/8,1/8,1]$.  We may assume that $\wtphi(B)$ is
empty.  To classify $B$, we look at the set $\wtphi(parent(B))$.
Say the translational and rotational parts of $parent(B)$ are
$[-1/2,3/2]^2$ and $[-1/8, 3/8, 1]$, respectively.
In this case $\wtphi(parent(B))$ contains any $g_0$ (and possibly $f_0$).
In any case, $g_0$ would be regarded as the closest feature in
$\wtphi(parent(B))$ because we use $\Sep'(C,f)$ for comparison.
Based on $g_0$, our algorithm would decide that $B$ is $\free$
when in fact $B$ is $\stuck$.

\end{document}